\documentclass[12pt]{article}
\usepackage{amssymb}
\usepackage{amsmath}
\usepackage{amsfonts}
\usepackage{graphicx}
\usepackage{epsfig}

\newcommand{\be}{\begin{eqnarray}}
\newcommand{\ee}{\end{eqnarray}}
\newcommand{\etal}{{\it et al.}}
\def\nue{{\nu_e}}
\def\anue{{\bar\nu_e}}
\def\numu{{\nu_{\mu}}}
\def\anumu{{\bar\nu_{\mu}}}
\def\nutau{{\nu_{\tau}}}
\def\anutau{{\bar\nu_{\tau}}}

\newcommand{\ms}{\Delta m^2_{21}}
\newcommand{\ma}{\Delta m^2_{31}}

\newcommand{\sss}{\sin^2 \theta_{12}}
\newcommand{\sch}{\sin^2 \theta_{13}}

\newcommand{\sig}{$3\sigma$}

\begin{document}

\title{\hspace{4.1in}\\
\hspace{4.1in}{\small OUTP 0611P}\bigskip\\
Probing neutrino oscillations from supernovae shock waves via the
IceCube detector}
\author{Sandhya Choubey$^{1,2}$\thanks{email: \tt sandhya@mri.ernet.in}~,
N. P. Harries$^1$\thanks{email: \tt n.harries1@physics.ox.ac.uk}~,
G.G. Ross$^1$\thanks{email: \tt g.ross1@physics.ox.ac.uk}~\\\\
$^1${\normalsize \it The Rudolf Peierls Centre for Theoretical Physics,}\\
{\normalsize \it University of Oxford, 1 Keble Road, Oxford, OX1 3NP, UK}\\
\\
$^2${\normalsize \it Harish-Chandra Research Institute,} \\
{\normalsize \it Chhatnag Road, Jhunsi, Allahabad  211 019, INDIA}}

\date{}
\maketitle

\begin{abstract}
The time dependent neutrino oscillation signals due to the passage
of a shock wave through the supernovae are analyzed for the case of
three active neutrinos and also for the case that there are two
additional sterile neutrinos. It is shown that, even without flavour
identification and energy measurement, detailed information about
the masses and mixing angles of the neutrinos may be obtained with a
detector with excellent time resolution such as IceCube. Such a
signal would also give important information about the nature of the
shock wave within the supernovae.
\end{abstract}

%%%%%%%%%%%%%%%%%%%%%%%%%%%
\section{Introduction}
%%%%%%%%%%%%%%%%%%%%%%

The possibility that, in addition to the three flavours of light
neutrinos required by the Standard Model, there are light sterile
neutrinos has been severely constrained by neutrino oscillation
experiments. However there is still the outstanding problem of
explaining the data from the Liquid Scintillator Neutrino Detector
(LSND) \cite{lsnd} which, if confirmed, might require the existence
of at least one extra light neutrino. The precision measurements at
LEP have shown that only three neutrinos with standard weak
interactions exist with mass less than $M_{Z}/2$ and therefore
additional neutrinos must be sterile \cite{sterileold}. However the
addition of a single sterile neutrino to explain the LSND anomaly is
not able to give a good fit to solar and KamLAND \cite{solar,kl2}
and atmospheric \cite{skatm,k2k} data together with the data from
the short-baseline experiments CHOOZ \cite{chooz} Bugey, CCFR84,
CDHS, KARMEN and NOMAD \cite{sbl}. While a 3+1 hierarchy in the four
neutrino case is preferred over a 2+2 hierarchy, neither is
acceptable \cite{analyses} and one is driven to consider the case of
two additional sterile neutrinos \cite{threeplustwo}. In this case a
3+2 mass hierarchy gives an acceptable fit, significantly better
than that found for the 3+1 case.

Pinning down the properties of additional sterile neutrinos is a
daunting task. Even if the LSND result is not confirmed there will
still be room for light sterile neutrinos mixing with the three
active states that terrestrial experiments will not be able to
exclude. Here we investigate the possibility of gaining significant
new information on this sector from the neutrinos coming from
supernovae. The reason this can give a significant signal is due to
the possibility of resonant conversion of the various neutrino
species (the Mikheyev-Smirnov-Wolfenstein (MSW) effect \cite{msw})
within the supernova. The MSW effect, being a resonant process, is
much more sensitive to small mixing angles than non resonant
oscillation phenomena. Feasibility studies of using neutrino signal
from a galactic supernova to determine the neutrino mass hierarchy
and constrain the mixing angle $\theta_{13}$ have been done before
in \cite{3nusnactive} for the standard case of three active
neutrinos and in \cite{4nusnsterile} for three active and one
sterile neutrino.

A further significant advantage of the supernovae signals for
neutrino oscillation has been appreciated in recent years
\cite{fullerfs}. This follows from the fact that in the supernova
explosion a shock wave propagates through the supernova. From the
theoretical simulations of supernovae it is believed that when the
core of the collapsing star reaches nuclear density, the collapse
rebounds forming a strong outward shock. This is stalled and then
regenerated by a neutrino driven wind. During this process it is
believed that as well as a forward shock a reverse shock could also
form \cite{raffeltforplusrev}. At the shock front the density
changes very rapidly so that the resonant transition may become
non-adiabatic as the shock wave passes
\cite{fullerfs,raffeltforplusrev}\footnote{While completing this
paper we received a paper \cite{turbulence} studying the effects of
turbulence on the shock wave. These effects have not be included
here but may be expected to broaden the structures presented
below.}. As a result one may have rapid changes in the active
neutrino luminosity produced by the supernova, very characteristic
signals which do not require knowledge of the overall luminosity.
These signals are observable by detectors such as IceCube
\footnote{For detailed discussion of the effect of the shock wave on
the signal in water Cerenkov detectors, we refer the readers to
\cite{fullerfs,oldfsonly,bargerfs} for effects of the forward shock
only and \cite{raffeltforplusrev,lisiforplusrev} for both forward
and reverse shocks.} capable of collecting high statistics with
excellent time resolution. IceCube will provide a $km^{3}$ neutrino
detector with a time resolution of 10ns \cite{icecube}. Although
designed to observe high energy neutrinos from astrophysical
sources, it has been realized that the lower energy neutrinos from
supernovae can be probed in IceCube through the detection of
additional photons in the background halo, caused by supernovae
neutrinos interacting with the ice \cite{raffelticecube}. IceCube
cannot measure the energy spectra
%or the flavour
of the supernova neutrinos, but because of its excellent time
resolution and high statistics, we find, for the reasons given
above, that it can be an excellent probe of a sterile neutrino
component to neutrino mass eigenstates. Moreover observation of such
signals in effect provide information on supernovae seismology and
can yield important information on the nature of the shock wave
within the supernovae.

For the case of oscillation between active (anti)neutrinos only,
detection of such effects requires a difference in the initial
properties of the neutrinos produced at the neutrinosphere. However
there is considerable uncertainty in this. Early studies
\cite{totani97} suggested that, while the luminosities of the three
antineutrino species should be quite close,
the average energy of the $%
\overline{\nu }_{e}$ should be about half
that of $\overline{\nu }_{\mu }$ and $%
\overline{\nu }_{\tau }$. More recently studies \cite{snnew}
including additional scattering processes within the neutrinosphere
have suggested that the difference in the average energies of the
$\anue$ and $\anumu/\anutau$ is quite small but that there is a
significant difference in their luminosities. We will investigate
the time dependent signals for active neutrino oscillation for both
these cases. For the case of oscillation to sterile neutrinos the
time dependent oscillation signals are not so sensitive to
uncertainties in the initial antineutrino processes because sterile
neutrinos produced by resonant oscillation will not be visible in
detectors and so oscillation to a sterile neutrino is observable
even with a detector having no capability to distinguish the flavour
or energy of active neutrinos.

In this paper we study the signals to be expected by IceCube both
for the case of just three active neutrinos and for the case that
there are two additional sterile neutrinos. To make the discussion
tractable we concentrate on the range of mass and mixing angle
parameters that provide an explanation of the LSND events while
being consistent together with all other neutrino oscillation data.
However our results indicate that such signals will also be
significant in constraining the sterile neutrino component even if
the LSND result is not confirmed.

In Section 2 we briefly review resonant conversion within the
supernova, the various possibilities for the neutrino spectra and
list the parameters used in the subsequent analysis. In Section 3 we
discuss the properties of the neutrinos at the neutrinosphere and
the nature of the shockwaves propagating through the supernova.
Section 4 contains the results of our analysis. We first review the
characteristics of the IceCube detector and then present its
detection rates for a \textquotedblleft standard\textquotedblright\
supernova for the case of three active neutrinos only and for the
case of three active neutrinos plus two sterile neutrinos for the
various possible neutrino mass hierarchy possibilities. Finally in
Section 5 we present our conclusions.

%%%%%%%%%%%%%%%%%%%%%%%%%%%%%%%%%%%%%%%%%%%%%%%%%%%%
\section{Resonant Conversion Within the Supernova}
%%%%%%%%%%%%%%%%%%%%%%%%%%%%%%%%%%%%%%%%%%%%%%%%%%%

\subsection{Three Active Neutrinos}

We start with the case of three active (anti)neutrino species. In
the basis of flavor eigenstates the evolution of (anti)neutrinos is
described by the effective Hamiltonian
%%%%%%%%%%%
\begin{equation}
{\cal H}=\frac{1}{2E}(U{\cal M}^{2}U^{\dag } + {\cal A})~,
\label{eq:schrod}
\end{equation}
%%%%%%%%%%%%%%%
where $U$ is a unitary matrix and is defined by $|\nu _{i}\rangle
=\sum_\alpha U_{i\alpha}|\nu _{\alpha }\rangle$. The vacuum mass
eigenstates are $\nu _{i}$ and the weak interaction eigenstates are
$\nu _{\alpha }$. ${\cal M}^{2}$ is the matrix of mass squares of
the mass eigenstates in vacuo, and ${\cal A}$ is the matter induced
mass matrix. For mixing between three active neutrinos

\begin{equation}
{\cal M}^{2}=Diag(m_{1}^{2},m_{2}^{2},m_{3}^{2})~,
\end{equation}

\begin{equation}
{\cal A}=Diag(A_{1},0,0)~,
\end{equation}

\begin{equation}
A_{1}=\pm \sqrt{2}G_{F} \rho N_A Y_{e}\times 2E ~, \label{eq:3ccmat}
\end{equation}%
where $i=1-3$ and $\alpha =e,\mu $ or $\tau $, $m_{i}^{2}$ is the
mass squared of the $i^{th}$ vacuum mass eigenstate, $G_{F}$ is the
Fermi coupling constant, $\rho $ is the density of ambient matter,
$N_A$ is the Avogadro's number, $Y_{e}$ is the electron fraction and
$E$ is the energy of the neutrino. $A_{1}$ is the matter potential
induced by charged current interactions of $\nu_{e}$ ($\anue$) with
electrons and the ``$+$'' (``$-$'') sign in Eq. (\ref{eq:3ccmat})
corresponds to neutrinos (antineutrinos). There is also an effective
mass squared induced by neutral current interactions of the
(anti)neutrinos with matter. However, this does not have a physical
effect because the neutral current interaction is universal for all
(anti)neutrino flavours, while oscillations depend on the difference
of mass squared and not on the absolute value. The mixing matrix $U$
is given in terms of mixing angles and CP-violating phases. If CP
conservation is assumed the mixing matrix takes
the form%
\begin{equation}
U=\prod_{0<B<A}\prod_{A=2}^{3}R^{BA}(\theta _{AB}) ~,
\end{equation}%
where $R^{AB}(\theta _{AB})$ are rotation matrices, representing a
rotation of $(\theta _{AB})$ in the AB plane. The mixing matrix for
three flavours is therefore given as
\begin{equation}
U=\left(
\begin{array}{ccc}
c_{12}c_{13} & s_{12}c_{13} & s_{13} \\
-s_{12}c_{23}-c_{12}s_{23}s_{13} & c_{12}c_{23}-s_{12}s_{23}s_{13} &
s_{23}c_{13} \\
s_{12}s_{23}-c_{12}c_{23}s_{13} & -c_{12}s_{23}-s_{12}c_{23}s_{13} &
c_{23}c_{13}
\end{array} \right)~,
\end{equation}%
where $c_{ij}\equiv \cos \theta _{ij}$ and $s_{ij}\equiv \sin \theta
_{ij}$.

\begin{figure}[t]
\begin{center}
\includegraphics[width=17cm]{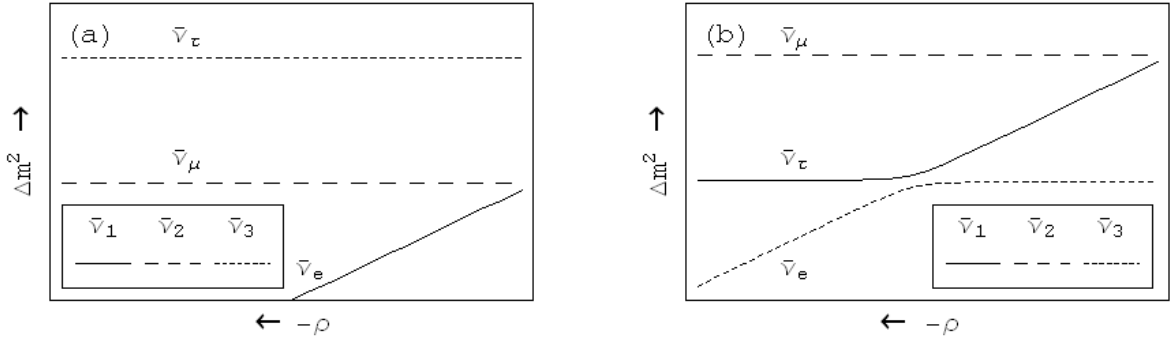}
\end{center}
\caption{The evolution with density of the mass eigenvalues
(squared) for the case of three active antineutrinos (a) for the
normal hierarchy ($\ma > 0$) and (b) for the inverse hierarchy ($\ma
< 0$).} \label{fig:3mass}
\end{figure}

The difference in mass squares of the mass eigenstates in matter can
be calculated by diagonalizing Eq. (\ref{eq:schrod}) and is shown in
Fig. \ref{fig:3mass} as a function of the matter density for the
antineutrinos\footnote{Since the matter potential $A$ is negative
for the antineutrinos, we show the evolution of the mass squares of
the mass eigenstates in Fig. \ref{fig:3mass} as a function of
$-\rho$, where $\rho$ is the matter density.}. Flavour oscillations
predominantly occur at the resonance densities, where the effective
mass difference between the two relevant mass eigenstates in matter
becomes minimum. If the resonances are far apart then each resonance
can be treated as an effective two neutrino problem, independent of
other resonances.
%The final oscillation
%probability can then be expressed in terms of
For two flavor oscillations the resonance condition is given by
\begin{equation}
A_1=\Delta m^{2}\cos 2\theta  ~, \label{eq:res}
\end{equation}%
where $\Delta m^2\equiv m_2^2 - m_1^2$ is the mass squared
difference and $\theta$ is the mixing angle between the two states
in vacuum. At the resonance a flip between the mass eigenstates is
possible due to the changing mass density. The ``flip probability''
between the
two mass eigenstates is given by%
\begin{equation}
P_{J}=\frac{\exp (-\gamma \sin ^{2}\theta )-\exp (-\gamma )}{1-\exp
(-\gamma )} \label{eq:jumpprob}
\end{equation}
\begin{equation}
\gamma =\pi \frac{\Delta m^{2}}{E}\left\vert \frac{d\ln
A_1}{dr}\right\vert _{r=r_{mva}}^{-1}
\end{equation}%
where $r_{mva}$ is the position of the maximum violation of
adiabaticity
($mva$) \cite{mva} and is defined as%
\begin{equation}
A_1(r_{mva})=\Delta m^{2}
\end{equation}
For a given set of values of $\Delta m^2$ and $\theta$, if the
density gradient is small enough so that $\gamma \gg 1$ and $\gamma
\sin^2\theta \gg 1$, the flip probability reduces to $P_{J}=\exp
(-\gamma \sin ^{2}\theta )\simeq 0$ and the resonance is called
adiabatic. However, if for the given $\Delta m^2$ and $\theta$, the
density gradient is large such that $\gamma \ll 1$, then
$P_{J}\simeq \cos ^{2}\theta$. In that case for very small mixing
angles $P_{J}\simeq 1$ and the resonance is completely
non-adiabatic. For all intermediate regions $P_J$ ranges between
$[0-1]$.

For three active neutrinos with their vacuum masses given by the
current experimental data and for a static supernova density
profile, the resonance condition is satisfied at two distinct widely
separated densities inside the supernova \cite{kuopanta}. The
resonance at the higher density is driven by $\ma$ and
$\theta_{13}$, while that at lower densities is driven mainly by
$\ms$ and $\theta_{12}$. Since the matter induced potential $A_1$ is
positive (negative) for neutrinos (antineutrinos), the resonance
condition given by Eq. (\ref{eq:res}) is satisfied only if $\Delta
m^2$ is positive (negative). Since $\ms$ (for $\theta_{12} < \pi/4$)
is known to be positive at a very high level of confidence from the
solar neutrino data, the lower resonance is therefore satisfied only
for the neutrino channel. However, $\ma$ could be either positive
(normal hierarchy) or negative (inverted hierarchy) and therefore
the higher resonance can occur either in the neutrino or the
antineutrino channel.
%This can be seen in Fig. \ref{fig:3mass}.
It can be shown that for the most plausible supernova density
profiles and the current allowed values of $\ms$ and $\theta_{12}$,
the flip probability at the lower resonance is zero and the
transition is completely adiabatic. However, depending on the value
of $\theta_{13}$ and the neutrino mass hierarchy, the flip
probability at the $\ma$ driven higher resonance may have any value
between [$0-1$]. We refer readers to \cite{3nusnactive} for detailed
discussion of the three generation oscillation probability in a
supernova environment with static density profiles.

Things get more involved when one considers the effect of shock
waves on the supernova density profiles. As we will discuss in the
following section, the effect of the shock is to cause very sharp
jumps in the density gradient. This results in the same $\Delta m^2$
producing multiple resonances which are relatively close together.
If we assume that the phase effects can be neglected even in this
case\footnote{This is a good approximation especially in our case
since we will be working with the IceCube detector which does not
have any energy sensitivity. This means that the detected supernova
events will be averaged over energy and hence phase effects
\cite{phase} will be further washed out.}, then the individual
resonances can be considered as independent two generation
resonances and the net flip probability $P_H$ can be expressed in
terms of the multiple flip probabilities $P_i$ as
\cite{kuopanta,oldfsonly} \be
\begin{pmatrix}
1-P_H & P_H\\
P_H & 1-P_H
\end{pmatrix}
= \prod_{i=1,n}
\begin{pmatrix}
1-P_i & P_i\\
P_i & 1-P_i
\end{pmatrix}
~, \label{eq:multiplePJ} \ee where $n$ is the number of resonances
occurring for the same $\Delta m^2$ due to the shock effect.

%%%%%%%%%%%%%%%%%%%%%%%%%%%%%%%%%%%%%%%%%%%%%%%%%%%%%%%%%%%%
\subsection{Inclusion of Two Sterile Neutrinos}
%%%%%%%%%%%%%%%%%%%%%%%%%%%%%%%%%%%%%%%%%%%%%%%%%%%

\begin{figure}[p]
\begin{center}
\includegraphics[width=17cm]{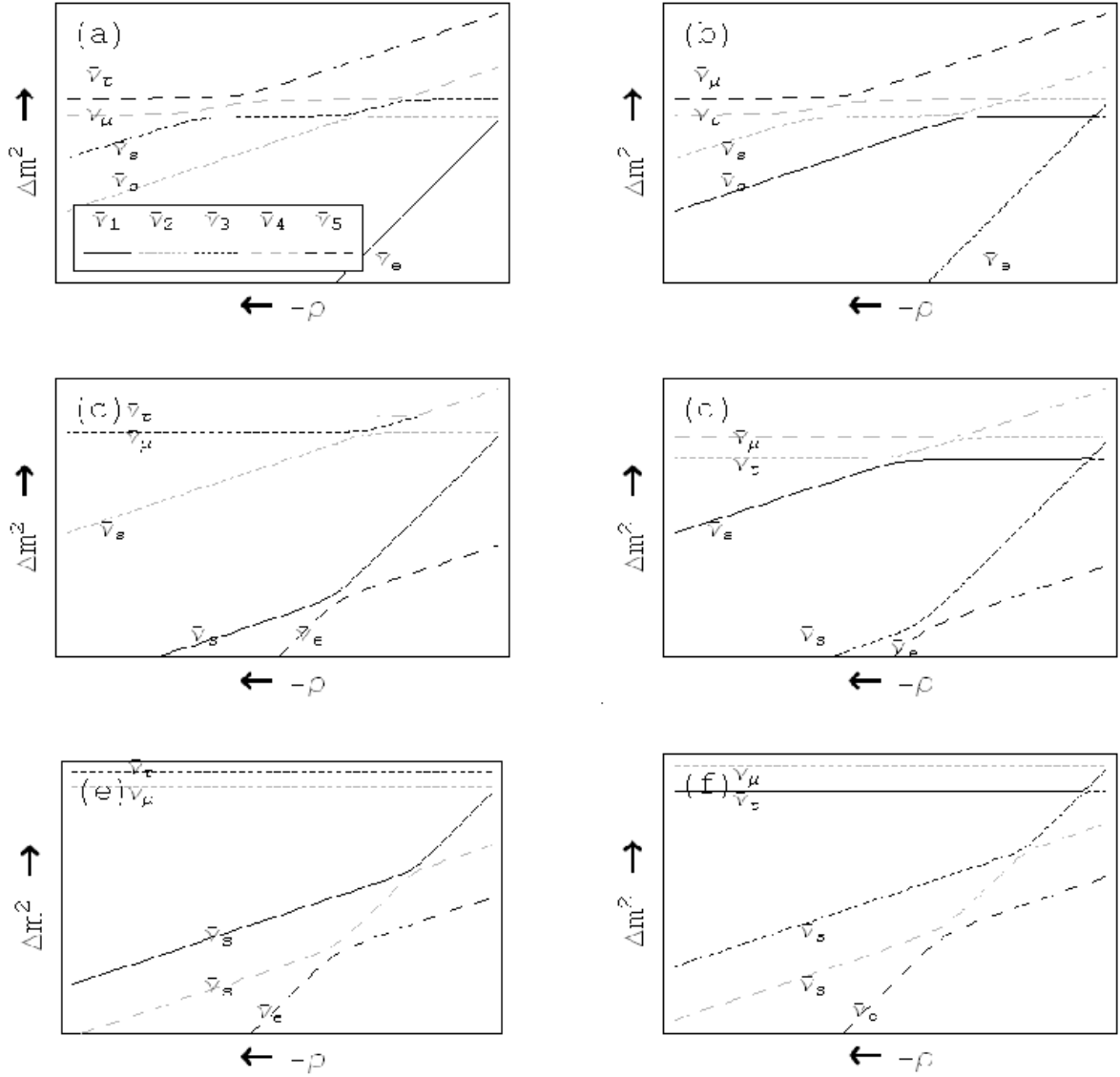}
\end{center}
\caption{The evolution with density of the mass eigenvalues
(squared) for the case of three active anti-neutrinos and two
sterile anti-neutrinos, for the mass hierarchies (a) N2+N3, (b)
N2+I3, (c) H2+N3, (d) H2+I3, (e) I2+N3 and (f) I2+I3. N3 and I3
correspond to the normal and inverse hierarchies in the active
neutrino sector. N2 corresponds to the normal hierarchy in the
sterile sector where $\Delta m_{51}^{2}>0$ and $\Delta
m_{41}^{2}>0$. H2 corresponds to the half hierarchy in the sterile
sector where either $\Delta m_{51}^{2}>0$ and $\Delta m_{41}^{2}<0$,
or $\Delta m_{51}^{2}<0$ and $\Delta m_{41}^{2}>0$. I2 corresponds
to the inverse hierarchy in the sterile sector where $\Delta
m_{51}^{2}<0$ and $\Delta m_{41}^{2}<0$.} \label{fig:3+2mass}
\end{figure}

For mixing between 3 active neutrinos and 2 sterile neutrinos the
form of Eq. (\ref{eq:schrod}) remains the same but now we have
\begin{equation}
U=\prod_{B>A}^{5}\prod_{A=1}^{4}R^{AB}
\end{equation}%
where $R^{AB}$ is a 5 by 5 rotation matrix about the AB plane,
\begin{equation}
{\cal M}^{2}=Diag(m_{1}^{2},m_{2}^{2},m_{3}^{2},m_{4}^{2},m_{5}^{2})
\end{equation}
\begin{equation}
{\cal A}=Diag(A_{1},0,0,A_{2},A_{2}) \label{eq:5mat}
\end{equation}
\begin{equation}
A_{1}=\pm \sqrt{2}G_{F} \rho N_A Y_{e}\times2E \label{eq:5ccmat}
\end{equation}
\begin{equation}
A_{2}=\pm \sqrt{2}G_{F}\rho N_A (1-Y_{e})\times E
\end{equation}%
where, as in Eq. (\ref{eq:3ccmat}), the ``$+$'' and ``$-$'' signs in
Eq. (\ref{eq:5ccmat}) corresponds to neutrinos and antineutrinos
respectively. As before,
$i=1-5$ and $\alpha $=e, $\mu $, $\tau $, $s_{1}$ or $s_{2}$, where $%
s_{1}$ and $s_{2}$ are two sterile neutrinos. $A_{2}=-A_{NC}$, where
$A_{NC}$ is the matter-induced potential of the neutral current
interaction of neutrinos/antineutrinos with matter. As discussed in
the previous section, all the three active neutrinos/antineutrinos
pick up an equal matter-induced potential $A_{NC}$ in matter and the
matter induced mass matrix is given by \be {\cal A}^\prime &=&
Diag(A_{1}+A_{NC},A_{NC},A_{NC},0,0)~,
\nonumber\\
&=& {\cal A} + A_{NC}{\cal{I}}~. \ee Since the sterile neutrinos do
not interact with matter, $A_{NC}$ cannot be completely factored out
from the mass squared matrix as in the three neutrino case and hence
can be recast as a negative matter potential for the sterile states.
The effective mass squared differences between the
neutrino/antineutrino eigenstates in matter therefore depend on both
$A_1$ and $A_2$.

The difference in mass squared can again be obtained by
diagonalizing the mass squared matrix in matter and the results for
the antineutrino channel are shown in Fig. \ref{fig:3+2mass} for the
different possible mass hierarchies of the five (anti)neutrino
system in vacuum. As discussed above, while we know that $\ms>0$,
sign of $\ma$ is presently unknown. The only experimental
information we have on $\Delta m^2_{41}$ and $\Delta m^2_{51}$ comes
from the data of short baseline experiments including LSND and their
sign are completely unknown as well. Therefore, in principle its
possible to have six types of mass hierarchy. For simplicity of
notation, we will
henceforth call them: \\
(a) N2+N3: $\ma>0$, $\Delta m^2_{41}>0$ and $\Delta m^2_{51}>0$, \\
(b) N2+I3: $\ma<0$, $\Delta m^2_{41}>0$ and $\Delta m^2_{51}>0$, \\
(c) H2+N3: $\ma>0$, $\Delta m^2_{41}>0$ and $\Delta m^2_{51}<0$, \\
(d) H2+I3: $\ma<0$, $\Delta m^2_{41}>0$ and $\Delta m^2_{51}<0$, \\
(e) I2+N3: $\ma>0$, $\Delta m^2_{41}<0$ and $\Delta m^2_{51}<0$, \\
(f) I2+I3: $\ma<0$, $\Delta m^2_{41}<0$ and $\Delta m^2_{51}<0$.

As discussed in the previous section the passage of the shock wave
causes each $\Delta m^{2}$ to have multiple resonances. The
additional resonances due to the inclusion of sterile neutrinos can
also be considered as independent, and the net flip probability is
given by equation \ref{eq:multiplePJ}.

%%%%%%%%%%%%%%%%%%%%%%%%%%%%%%%%%%%%%%%
\subsection{Experimental Bounds}
%%%%%%%%%%%%%%%%%%%%%%%%%%%%%%%%%

Here we very briefly review the current knowledge we have about the
masses and mixing angles of the active and sterile neutrinos from
all available neutrino oscillation data.

%Assuming CP
%conservation the the mixing matrix takes the form
%
%\begin{equation}
%U=\prod_{B>A}^{5}\prod_{A=1}^{4}R^{AB}
%\end{equation}%
%where $R^{AB}$ is a 5 by 5 rotation matrix about the AB plane.

%%%%%%%%%%%%%%%%%%%%%%%%%%%%%%%
\subsubsection{Solar Data}
%%%%%%%%%%%%%%%%%%%%%%%%%

The combined analysis of world data on solar neutrinos \cite{solar}
and the KamLAND reactor data \cite{kl2} has established the
so-called Large Mixing Angle (LMA) solution as the solution to the
solar neutrino anomaly with the current best-fit parameters
\cite{solanalysis} $\Delta m_\odot^2 \equiv \ms = 8.0\times 10^{-5}$
eV$^2$ and $\sin^2\theta_\odot \equiv \sss = 0.31$ and the \sig{}
allowed range given by \cite{solanalysis,screv}
\begin{eqnarray}
0.25 < & \sin^{2}\theta_{12}& < 0.39 \\
7.2 \times 10^{-5} {\rm eV}^{2}<& \Delta m^{2}_{21}& < 9.2\times
10^{-5}{\rm eV}^{2}
\end{eqnarray}

%%%%%%%%%%%%%%%%%%%%%%%%%%%%%%%%%%%%%%%%%
\subsubsection{Atmospheric Data}
%%%%%%%%%%%%%%%%%%%%%%%%%

The zenith angle dependent event spectrum of the atmospheric
neutrino data from Super-Kamiokande \cite{skatm} and the data from
the long baseline K2K experiment \cite{k2k} can be best explained in
terms of almost pure $\numu-\nutau$ oscillations with best-fit
parameters $\Delta m_{\rm atm}^2 \equiv |\ma| = 2.1\times 10^{-3}$
eV$^2$ and $\sin^22\theta_{\rm atm} \equiv \sin^22\theta_{23} = 1$
and the \sig{} allowed range given by \cite{skatm}
\begin{eqnarray}
\sin^{2}2\theta_{23}&>&0.9 \\
1.3\times 10^{-3}{\rm eV}^{2}<&\Delta m^{2}_{31}& < 4.2\times
10^{-3}{\rm eV}^{2}
\end{eqnarray}

%%%%%%%%%%%%%%%%%%%%%%%%%%%%%%%%%%%%%
\subsubsection{CHOOZ Reactor Data}
%%%%%%%%%%%%%%%%%%%%%%%%%

The upper limit on the mixing angle $\theta_{13}$ is mainly
determined by the reactor neutrino experiments CHOOZ and Palo Verde
\cite{chooz}. Data from these experiments when combined with the
solar and atmospheric neutrino data gives at $3\sigma$ the bound
\cite{screv,global}. \be \sch < 0.044 \ee\label{eq:CHOOZ}

%%%%%%%%%%%%%%%%%%%%%%%%%%%%%%%%%%%%%%%%%%%%
\subsubsection{Data from Short Baseline Experiments}
%%%%%%%%%%%%%%%%%%%%%%%%%%%%%

A combined analysis of Bugey, CHOOZ, CCFR84, CDHS, KARMEN, NOMAD,
and LSND for the case of a 3+2 mass hierarchy gives two possible
solutions \cite{threeplustwo}. These are shown in Table
\ref{tab:accdata}. Column 2 of Table \ref{tab:accdata} shows the
global best-fit, while column 3 gives the solution if all neutrino
masses were restricted to lie in the sub-eV range \footnote{Note
that for the best-fit solution shown in column 2, the sum of
neutrino masses are in conflict with the current cosmological bounds
(see for instance \cite{cosmobounds} and reference therein).
However, we still use them in this paper as an illustrative example
for the case where the sterile neutrinos are non-degenerate.}. Note
that the values of $\theta_{34}$, $\theta_{35}$ and $\theta_{45}$
are not constrained by any experiments to date. However in what
follows, we shall assume that all concerned mixing angles (including
$\theta_{13}$) are large enough so that all the transition
probabilities are adiabatic for the static density profile of the
supernova in absence of shock effects. We shall briefly discuss in
section 4.4 the effect of reducing the mixing angles and hence the
degree of adiabaticity on the resultant supernova neutrino signal.

%TCIMACRO{\TeXButton{B}{\begin{table}[tbp] \centering}}%
%BeginExpansion
\begin{table}[tbp] \centering%
%EndExpansion
\begin{tabular}{|c|c|c|}
\hline Parameter & Best fit & Best fit in sub-eV\\ \hline
$\Delta m^2_{41}$ & 0.92 eV$^{2}$ & 0.46 eV$^{2}$ \\
$\Delta m^2_{51}$ & 22 eV$^{2}$ & 0.89 eV$^{2}$ \\
$U_{e4}$ & 0.121 & 0.090 \\
$U_{\mu 4}$ & 0.204 & 0.226 \\
$U_{e5}$ & 0.036 & 0.125 \\
$U_{\mu 5}$ & 0.224 & 0.160 \\ \hline
\end{tabular}%
\caption{The two best fit solutions for the short baseline
experiments taken from \cite{threeplustwo}.}\label{tab:accdata}%
%TCIMACRO{\TeXButton{E}{\end{table}}}%
%BeginExpansion
\end{table}%
%EndExpansion

%%%%%%%%%%%%%%%%%%%%%%%%%%%%%%%%%%%%%%%%%%%%%%
\section{Supernova Neutrinos and Shock waves}
%%%%%%%%%%%%%%%%%%%%%

%%%%%%%%%%%%%%%%%%%%%%%%%%%%%
\subsection{Time Dependent Luminosity and Energy Spectra.\label%
{energyspectra}}
%%%%%%%%%%%%%%%%%%%%%%%%%

About $3\times 10^{53}$ ergs of energy is released in a type-II
supernova, 99\% of which is in the form of neutrinos. These
neutrinos drift out from the dense core of the proto-neutron star
and beyond a certain radius determined by the energy of the
neutrino, travel freely and escape. This radius of last scattering
which characterizes the energy distribution of the neutrinos is
generally called the ``neutrinosphere''. Neutrinos are mainly
released in two phases. A short ``neutrinonization burst'' of pure
$\nue$ is produced by electron capture on protons when shock wave
crosses the neutrinosphere. Subsequently the majority of the
neutrinos are produced as $\nu-\bar\nu$ pairs of all three flavours
over a period of 10-20 seconds as the proto-neutron star cools.

\begin{figure}[h]
\begin{center}
\includegraphics[width=7cm]{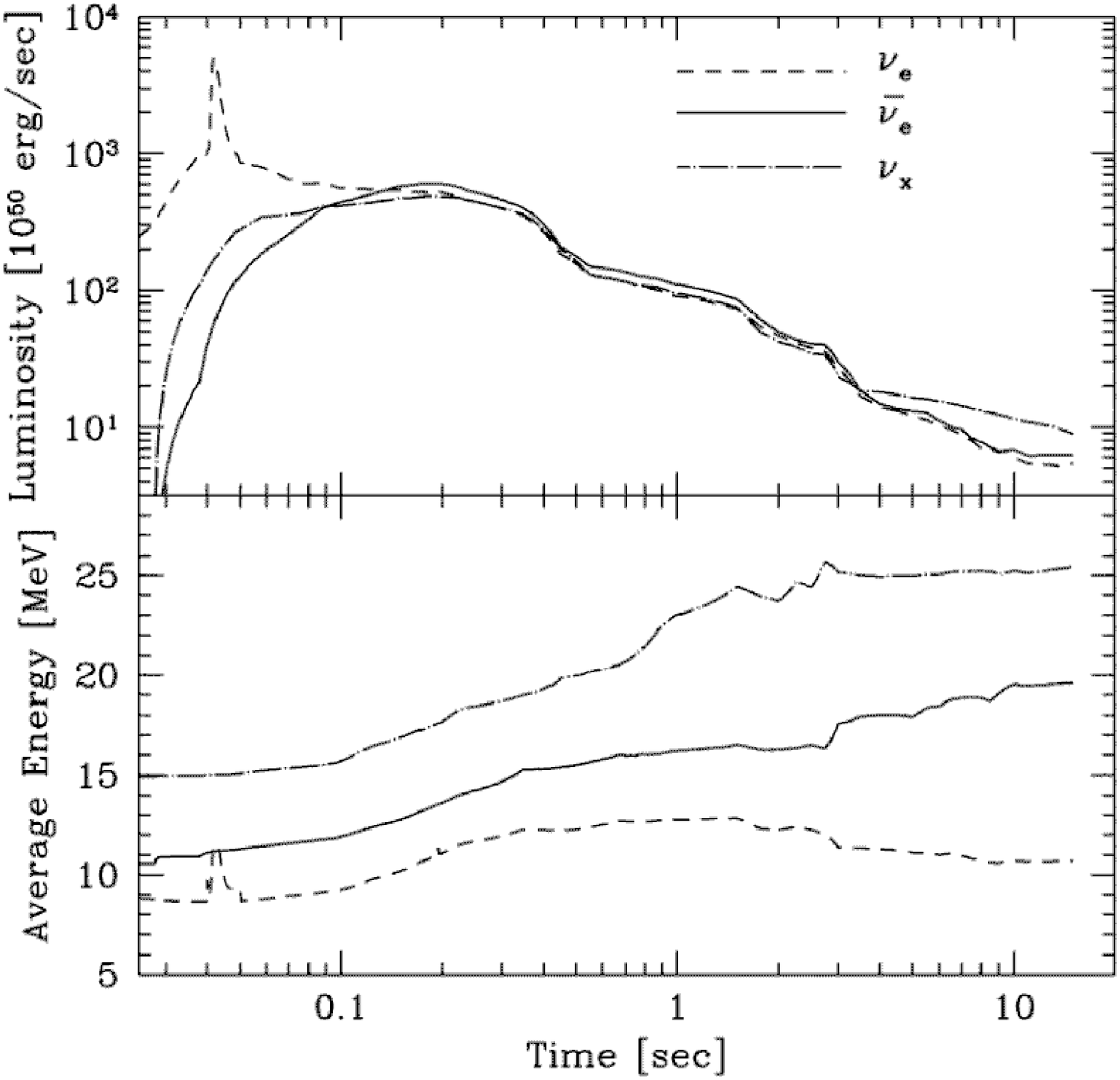}
\end{center}
\caption{The time evolution of the luminosity and average energy as
calculated by the Lawrence Livermore group.}
\label{fig:lumandavenergy}
\end{figure}

Even if all the six species of neutrinos are dominantly produced by
the same mechanism, their luminosities and energies at their
respective neutrinospheres can differ due the difference in their
degree of interaction with matter. If the effect of weak magnetism
and muon production inside the supernova core is neglected, the
spectra of $\nu_{\mu }$, $\anumu$ $\nu_{\tau }$ and $\anutau$ are
approximately equal, and so they are often grouped together and
collectively denoted as $\nu_{x}$. An accurate prediction of the
time dependent energy spectra of the emitted neutrinos requires a
full simulation including all significant neutrino interactions. To
date only the Lawrence Livermore group \cite{totani97} has published
detailed results of the energy spectra of the neutrinos over the
full duration of the supernova, in a simulation with a successful
explosion. The resulting luminosity and average energies of the
neutrinos taken from \cite{totani97} are shown in Fig.
\ref{fig:lumandavenergy}. It may be seen that the luminosities of
all the six types of neutrinos are very similar (apart from the
$\nue$ neutronization burst),
%$\nu_{x}$ (and $\bar\nu_x$) are very close to
%that of $\anue$
but the average energies of $\nu_{x}$ are considerably larger than
that of $\nue$ and $\anue$ throughout the duration of the supernova.
Since neutrino oscillations effectively flip the energy spectra of
the $\nue$ and/or $\anue$ with that of the $\nu_x$ and since the
extent of this flip depends on the value of the mixing angle and the
neutrino mass hierarchy, this difference in the initial energy
spectra of the different (anti)neutrino species has been used in
previous studies in the context of three \cite{3nusnactive} and four
\cite{4nusnsterile} neutrino oscillations to put bounds on the
neutrino mass spectrum and mixing angles by exploiting the energy
dependence of the cross sections relevant to the detectors on earth.
However, more recent work \cite{snnew} on neutrino transport inside
the supernova has cast doubt on the magnitude of this effect. When
all neutrino scattering processes, some of which were not included
in the Livermore study, are included, the average energies of the
different (anti)neutrino flavors, particularly $\anue$ and $\nu_x$
become very similar. The condition of equipartition of luminosity
between the different species also breaks down and in some
simulations one might find the $\anue$ luminosity to be almost
double that of $\nu_x$.

In what follows, we shall use the time dependence of the luminosity
as given by the Livermore group in our estimates of neutrino
oscillation effects as this should not be so sensitive to the
additional scattering processes not included in the Livermore
analysis. However when considering the oscillations among the three
active neutrinos we will estimate the uncertainties by computing the
expected signal both for the Livermore luminosities and energy
distributions and for the case where the energy distributions are
equal but the luminosities differ in the manner just discussed. For
the case of oscillation into sterile neutrinos the results are
relatively insensitive to these differences as the signal does not
depend on a difference between $\overline{\nu }_{x}$ and
$\overline{\nu}_{e}$.

%%%%%%%%%%%%%%%%%%%%%%%%%%%%%%%%%%%%%%%%%%
\subsection{Supernovae Shock Wave(s)}
%%%%%%%%%%%%%%%%%%%%%%%%%%%%%%%%%%

\begin{figure}[h]
\begin{center}
\includegraphics[width=17cm]{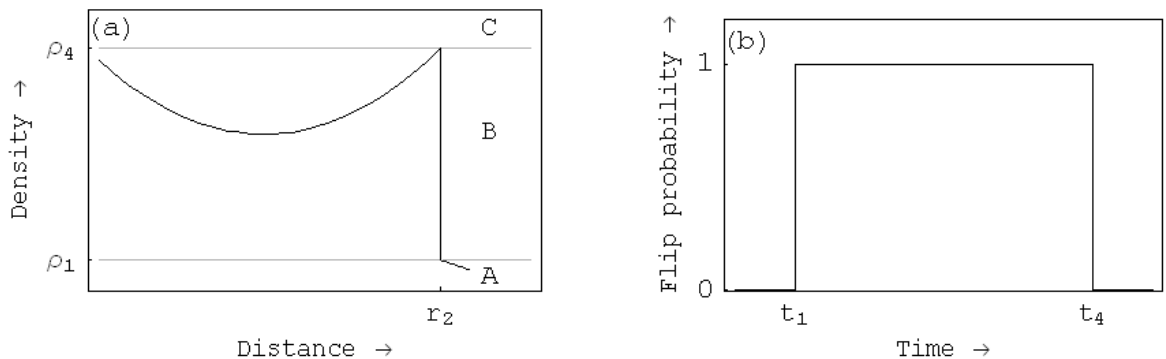}
\end{center}
\caption{(a) The density profile of the forward shock as a function
of the distance from the core of the supernova. (b) The resulting
flip probability as a function of time. } \label{fig:forward}
\end{figure}

\begin{figure}[h]
\begin{center}
\includegraphics[width=17cm]{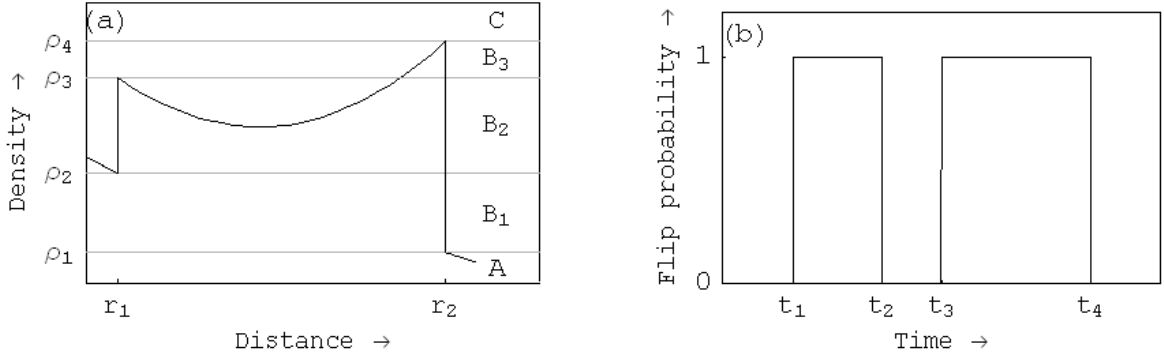}
\end{center}
\caption{(a) The density profile of the forward and reverse shock as
a function of the distance from the core of the supernova. (b) The
resulting flip probability as a function of time. }
\label{fig:forwardandback}
\end{figure}

In this analysis we are particularly interested in the shock waves
that form within the supernovae.
%As discussed above
It is believed that when the core of the collapsing star reaches
nuclear density, the collapse rebounds forming a strong outward
shock. This is stalled and then regenerated by a neutrino driven
wind. During this process both a forward and a reverse shock may
form \cite{raffeltforplusrev}. In order to get a realistic estimate
of the density profile of the shock wave it is necessary to use
results from numerical simulations. However, since detailed
numerical results from such simulations are not available to us, we
will use a simplified profile for the shockwave used in
\cite{oldfsonly,lisiforplusrev}. In order to exhibit the impact of
both the forward and reverse shock on the resultant neutrino signal
in earth-bound detectors, we will consider scenarios where we (i)
neglect the effect of the shock wave, (ii) consider the effect of
the forward shock alone and (iii) consider both the forward and
reverse shock in calculating the oscillation probabilities. The snap
shot of the density profile at a certain post-bounce time $t$ for
the case of a forward shock only is shown in Fig. \ref{fig:forward}a
and the case of both a forward and reverse shock is shown in Fig.
\ref{fig:forwardandback}a. The shape of the shockwave remains
roughly constant in time but the height varies as the shockwave
propagates through the supernovae (c.f.Fig 1 \cite{oldfsonly}). The
effect of the shock wave on the neutrino oscillation probabilities
has been discussed before in \cite{fullerfs,oldfsonly} for the
forward shock and in \cite{raffeltforplusrev,lisiforplusrev} for the
forward and reverse shock. We will briefly review it again in the
next subsections for completeness.

%%%%%%%%%%%%%%%%%%%%%%%%%%%%%%%%%%%%%%%%%%%%%
\subsection{Effect of the Shock Wave}
%%%%%%%%%%%%%%%%%%%%%%%%%%%%%%%%%%%%%%%

%%%%%%%%%%%%%%%%%%%%%%%%%%%%%%%%%%%%%
\subsubsection{ Forward Shock Only}
%%%%%%%%%%%%%%%%%%%%%%%%%%%

We will consider here the effect of the shockwave on the neutrinos
emerging from the supernova in the case where all the relevant
mixing angles
%the undetermined mixing angles
%$\theta_{13}$,
%$\theta_{34}$, $\theta_{35}$ and $\theta _{45}$
are sufficiently large so that the resonance for the static density
profile in the absence of the shockwave is completely adiabatic.

The density structure shown in Fig. \ref{fig:forward}a exhibits
three separate regions. Initially, for times $t<t_{1},$ the resonant
density $\rho _{R}$  $(= (\Delta m^2\cos2\theta)/(2\sqrt{2}G_F N_A
Y_e E))$ lies below the density at which the shock wave appears,
$\rho _{R}<\rho_{1}$, i.e. in region A. The resonance crossing is
adiabatic to a good approximation in this region and therefore the
flip probability vanishes for $t<t_{1}$ as shown in Fig.
\ref{fig:forward}b. As the shockwave moves out $\rho _{1}$ decreases
and for times $t_{1}<t<t_{4} $ the resonant density lies in region
$B$, $\rho _{1}<\rho _{R}<\rho _{4}.$ Due to the large density
gradient at the shock front, in this region the flip probability
increases to $\cos ^{2}\theta \simeq 1$ (c.f. eq(\ref{eq:jumpprob}))
and therefore the resonance becomes non-adiabatic in the period
$t_{1}<t<t_{4}$ as shown in Fig. \ref{fig:forward}b. As may be seen
from the figure, the neutrino pass through two adiabatic resonances
and one non-adiabatic resonance. The flip probability passing
through each adiabatic resonances is approximately zero and
therefore the total flip probability is equal to the flip
probability passing through the non-adiabatic resonance alone.
Finally in region C the shockwave has passed through the resonant
density so that $\rho _{4}<\rho _{R}$ and the resonance is
adiabatic, therefore the flip probability is approximately zero
again for $t>t_{4}$.

%%%%%%%%%%%%%%%%%%%%%%%%%%%%%%%%%%%%%%%%%%
\subsubsection{\label{sec:shocks}Forward and Reverse Shock}
%%%%%%%%%%%%%%%%%%%%%%%%%%%%%%%%%%%%%%%

\begin{figure}[h]
\begin{center}
\includegraphics[width=10cm]{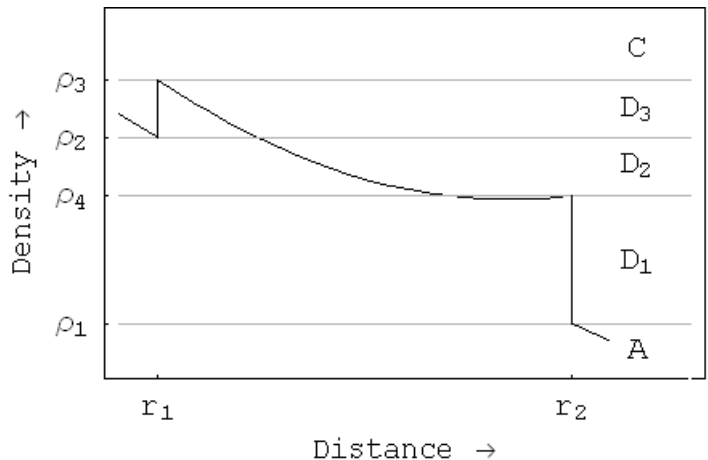}
\end{center}
\caption{An alternate density profile giving the same flip
probability as shown in figure \protect\ref{fig:forwardandback}.}
\label{fig:shockregion3}
\end{figure}

The situation is illustrated in Fig \ref{fig:forwardandback}. For
the density in the appropriate range there will now be two
non-adiabatic resonances, corresponding to the points $r_{1}$ and
$r_{2}$. As before at early times $t<t_{1},$ the resonant density
$\rho _{R}$ lies below the density at which the shock wave appears,
$\rho _{R}<\rho _{1},$ i.e. in region $A.$ As before, to a good
approximation in this region the resonance crossing is adiabatic and
so the flip probability vanishes for $t<t_{1}$ as is shown in Fig.
\ref{fig:forwardandback}b. As the shockwave moves out $\rho _{1}$
decreases and for times $t_{1}<t<t_{2}$ the resonant density lies in region $%
B_{1}$ such that $\rho _{1}<\rho _{R}<\rho _{2}$. Thus in region
$B_{1}$ only the forward shock has reached the resonant density and
is therefore equivalent to region B in the forward shock only case.
At later times $t_{2}<t<t_{3}$ the resonant density lies in region
$B_{2}$ where $\rho _{2}<\rho _{R}<\rho _{3}.$ In this region the
neutrino passes through the resonant density at both the forward and
reverse shock and both are non-adiabatic resonances. Therefore the
mass eigenstates are flipped at the reverse shock and then flipped
back at the forward shock.
%making the resulting total flip probability zero.
When the mixing angle is non-zero the flip probability at the
forward and reverse shocks is $\cos^{2}\theta $ and the total
flip probability is $\frac{1}{2}%
\sin ^{2}2\theta $. Between $t_{3}<t<t_{4}$, the shock profile
corresponds to region $B_3$ where $\rho _{3}<\rho _{R}<\rho _{4}$
and the neutrinos cross their resonance density through the forward
shock only. Hence the flip probability again increases suddenly to
$\cos^2\theta$. For $t> t_4$, we are in region $C$ where
$\rho_{4}<\rho_{R}$ and the shockwave has passed through the
resonant density. The resonance is therefore again adiabatic. The
total flip probability of the neutrinos for the case where we have
both forward and reverse shock has the form of a double peak as is
shown in Fig. \ref{fig:forwardandback}. The same structure of the
flip probability can be caused by a density profile where the two
non-adiabatic resonances which are not "overlapping", such as Fig.
\ref{fig:shockregion3}, where the flip probability is only non-zero
in regions $D_1$ and $D_3$. However this also requires that there
exists a forward and reverse shock.

%%%%%%%%%%%%%%%%%%%%%%%%%%%%%%%%%%%%%%%%%%%%%%%%%%%%%%%%%%%%%%
\section{Probing Supernovae Neutrino Oscillations in IceCube}
%%%%%%%%%%%%%%%%%%%%%%%%%%%%%%%%

In this Section we present the neutrino induced signal expected in
the IceCube detector for a galactic supernova event. We first give
the results for the case of three active neutrinos and then take up
the case where two additional sterile neutrinos are added to the
neutrino mass spectrum. We first turn to the discussion of the
properties of the IceCube detector.

%%%%%%%%%%%%%%%%%%%%%%%%%%%%%%%%%%%%%%%%%%%%
\subsection{IceCube as a Supernova Detector}
%%%%%%%%%%%%%%%

\begin{table}
\begin{center}
\begin{tabular}{|c|c|c|c|c|c|}
\hline Model & $\langle E^0_{\nue} \rangle$ & $\langle E^0_{\anue}
\rangle$  & $\langle E^0_{\nu_x} \rangle$ &
$\frac{\Phi^0_{\nue}}{\Phi^0_{\nu_x}}$ &
$\frac{\Phi^0_{\anue}}{\Phi^0_{\nu_x}}$\cr \hline LL & 12 & 15 & 24
& 2.0 & 1.6 \cr G1 & 12 & 15 & 18 & 0.8 & 0.8 \cr G2 & 12 & 15 & 15
& 0.5 & 0.5 \cr \hline
\end{tabular}%
\end{center}
\caption{\label{tab:enlmodel} The average energies and total fluxes
characterising the primary neutrino spectra produced inside the
supernova. The numbers obtained in the Lawrence Livermore
simulations are denoted as LL, while those obtained by the Garching
group are denoted as G1 and G2. }
\end{table}

IceCube is a future $km^{3}$  neutrino telescope, under construction
in Antartica \cite{icecube}. On completion, it will contain 4800
optical modules deployed into the ice which will detect the Cerenkov
photons. Though designed to observe ultra high energy neutrinos,
IceCube can detect a supernova through its neutrinos by detecting
Cerenkov light from $e^{+}$ produced by $\anue$ capture on protons.
Each event cannot be distinguished in the detector, but the addition
to the background photon halo coming from the Cerenkov radiation
associated with the produced positron can be measured. The number of
additional photons is
\begin{equation}
N_{det}=\frac{47.75}{MeV m^{-3}}f_{Ch}f_{abs}f_{OM}\rho\frac{1}{4\pi
D^{2}} \int_{t_{min}}^{t_{max}}dt\int_{0}^{\infty }dE\sigma
(E)EF(E,t)
\end{equation}
where $f_{Ch}$, $f_{abs}$ and $f_{OM}$ are fudge factors of order 1
and are given in the appendix, $\rho=6.18\times10^{25}$ m$^{-3}$ is
the density of targets in ice, $\sigma$ (in cm$^2$) is the cross
section for $\bar{\nu}_{e}p\rightarrow ne^{+}$, $E$ (in MeV) is
energy, D (in cm$^2$) is the distance to the supernova, t (in s) is
time, F (in MeV$^{-1}$ s$^{-1}$) is the flux of anti-electron
neutrinos. The initial spectra of neutrino species $\nu_\alpha$ from
a supernova is parameterized as \cite{spectra}
\begin{equation}
F^{0}(E,t)=\frac{\Phi(t)}{\langle E \rangle(t) } \frac{(\alpha(t)+1
)^{\alpha(t)+1}}{\Gamma (\alpha(t) +1)} \left( \frac{E}{\langle
E\rangle(t) }\right) ^{\alpha(t) }\exp \left( -(\alpha(t)
+1)\frac{E}{\langle E \rangle(t) }\right) \label{eq:flux}
\end{equation}
where $\langle E\rangle $ and $\Phi$ are the average energy and
total number flux
%of the neutrino species $\alpha$
and $\alpha$ is a dimensionless parameter which typically takes the
values 2.5-5. For the results presented in this paper, we have
assumed $\alpha_{\bar e} = 3 $ and $\alpha_x = 4 $. In order to
compare the impact of the uncertainties on the average energies and
fluxes of the neutrinos obtained in different supernova computer
simulations, we will present our results using supernova neutrino
parameters given by both the Lawrence Livermore and Garching groups.
Specifically, we consider three cases shown in Table
\ref{tab:enlmodel} \cite{raffeltforplusrev}.

The neutrino flux in the detector is
\begin{equation}
F_{\beta }=\sum_{\alpha }F_{\alpha }^{0}P_{\alpha \beta }
\label{eq:fluxprob}
\end{equation}
where
\begin{equation}
P_{\alpha \beta }=\sum_{i}P_{\alpha i}^{m}P_{i\beta }^{\oplus }
\label{eq:flipprob}
\end{equation}
\begin{equation}
P_{\alpha i}^{m}=\sum_{j}|U_{\alpha j}^{m}|^{2}P_{i j}
\end{equation}
\begin{equation}
P_{i j}=|\langle \nu _{i}|\nu _{j}^{m}\rangle |^{2}
\end{equation}

$P_{\alpha i}^{m}$ is the probability that a $|\nu _{\alpha }\rangle
$ produced inside the supernova emerges in the $i^{th}$ mass
eigenstate, $|\nu _{i}\rangle $. $U_{\alpha j}^{m}$ is the mixing
matrix at the point of production, the neutrinosphere. $P_{i j}$ is
the probability that a $|\nu _{j}^{m}\rangle $ mass eigenstate in
matter appears as $|\nu _{i}\rangle $ mass eigenstate in a vacuum,
this is known as the flip probability. $P_{i\beta }^{\oplus }$ is
the probability of detecting the $|\nu _{i}\rangle $ mass eigenstate
in the $|\nu _{\beta }\rangle $ weak interaction eigenstate. In this
paper, we do not take into account Earth-matter effects and
therefore
\begin{equation}
P_{i\beta }^{\oplus }=|U_{i\beta }|^{2}
\end{equation}

%%%%%%%%%%%%%%%%%%%%%%%%%%%%%%%%%%%%%%%%%%%
\subsection{IceCube Background Signal}
%%%%%%%%%%%%%%%%%%%%%%%%%%%%%%

IceCube has an irreducible background photon halo with energies
comparable to those produced by the supernova neutrinos. In the
analysis of a supernova signal the additional number of photons need
to be distinguished from this background. The mean number of
detected photons is
\begin{equation}
\bar{n_{i}}=N_{M}\nu_{1pe}t_{bin}+N_{det}
\end{equation}
where $N_{M}$ is the number of optical modules, $\nu _{1pe}$ is the
mean background detection rate per optical module, $t_{bin}$ is
length of the time bin and $N_{det}$ is the signal. Using Poisson
statistics the fluctuations in the background photon halo is given
by $\sqrt{\bar{n_{i}}}$. The background rate is $500Hz$ per optical
module and IceCube would have 4800 optical modules. Thus for
time bin of 100ms the random fluctuations are expected to be $%
\sigma =\sqrt{2.4\times 10^{5}+N_{det}}$.

In addition there is an error due to the uncertainty in the time
resolution, $\delta t$, of the detector giving
\begin{eqnarray}
\Delta n_{i}^{+} &=&\frac{\delta t}{t_{bin}}(n_{i+1}+n_{i-1}) \\
\Delta n_{i}^{-} &=&\frac{\delta t}{t_{bin}}2n_{i}.
\end{eqnarray}%
where $t_{bin}$ is the size of each time bin and we take a detector
time resolution of $\delta t=10$ nsec for our estimate of the
uncertainties. In our analysis we will use as
an upper and lower estimate of the true event number the values $%
n_{i}+\Delta n_{i}^{+}+\sigma $ and $n_{i}-\Delta n_{i}^{-}-\sigma $
respectively, where $n_{i}$ is the number detected in the $i^{th}$
bin. In our numerical plots we will show only the additional photons
produced by the supernova neutrinos, but for the fluctuations we
will include all errors as discussed above.

%%%%%%%%%%%%%%%%%%%%%%%%%%
\subsection{Results}
%%%%%%%%%%%%%

%%%%%%%%%%%%%%%%%%%%%%%%%%%%%%%%%%%%%%%%%%
\subsubsection{Three Active Neutrinos}
%%%%%%%%%%%%%%%%%%%%%%%%%%

We first consider the case of a \textquotedblleft
standard\textquotedblright\ supernova at 10 kpc from earth with
average energy and luminosity for the neutrinos taken from the
Lawrence Livermore group simulation. The initial spectrum of
neutrinos used is that given in Eq. (\ref{eq:flux}).
%with
%$\alpha =4$ for $\bar{\nu}_{e}$ and $\alpha =3$ for $%
%\bar{\nu}_{x}$.
At the neutrinosphere the matter-induced potential is much greater
than the vacuum mass squared differences; for anti-neutrinos this is
large and negative. Therefore $\bar{\nu}_{e}$ is approximately equal
to the mass eigenstate with the lowest mass squared. In the normal
hierarchy this is $\bar{\nu}_{1}$ and in the inverted hierarchy this
is $\bar{\nu}_{3}$. Fig. \ref{fig:3mass} shows the evolution with
density of the mass squared of the anti-neutrino mass eigenstates
for the normal and inverted hierarchies. From this one may readily
see if there is a level crossing and determine the $\bar\nu_{e}$
flux in the detector using Eqs. (\ref{eq:fluxprob}) and
(\ref{eq:flipprob}). The relevant expressions for the probabilities
are given in Table \ref{tab:jumpprob3}. The resonant oscillation
effects for anti-neutrinos occur only if there is an inverted mass
hierarchy. In this case the flux of electron anti neutrinos in the
detector is given by
\begin{equation}
F_{e}=F_{x}^{0}+P_{ee}(F_{e}^{0}-F_{x}^{0}) \label{eq:flux1}
\end{equation}%
where%
\begin{equation}
P_{ee}=P_{13}|U_{e1}|^{2}+(1-P_{13})|U_{e3}|^{2}
\end{equation}%

\begin{table}[h]
\begin{centering}

\begin{tabular}{|c|c|c|c|}\hline
  Mass hierarchy & i & $P^{m}_{ei}$ & $P^{m}_{xi}$ \\
  \hline
  \hline
  Normal & 1 & 1 & 0 \\
         & 2 & 0 & 1 \\
         & 3 & 0 & 1 \\
  \hline
  Inverted  & 1 & $P_{13}$ & $1-P_{13}$ \\
            & 2 & 0 & 1 \\
            & 3 & $1-P_{13}$ & $P_{13}$ \\ \hline
\end{tabular}
\caption{\label{tab:jumpprob3} The probabilities $P_{ei}^m$ and
$P_{xi}^m$ for three active neutrinos, where $P^{m}_{\alpha i}$ is
given by Eq. (\ref{eq:flipprob}) and $P_{ij}$ is the flip
probability at the resonance between the $\nu_{i}$ and $\nu_{j}$
mass eigenstates. Only $P_{13}$ appears in the expression for the
probabilities since $P_{12}=0$ for the current values of $\ms$ and
$\sss$.}
\end{centering}
\end{table}

\begin{figure}[h]
\begin{center}
\hglue -2.5cm
\includegraphics[width=17cm]{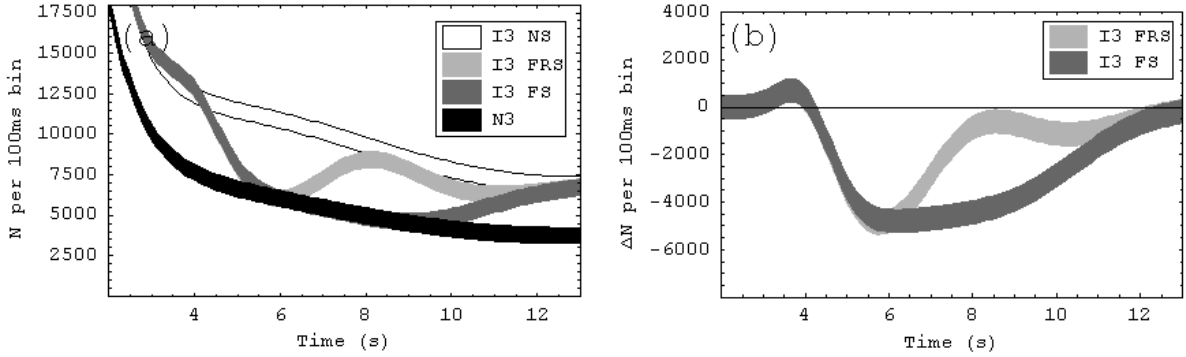}
\end{center}
\caption{\label{fig:3active} The left hand panel shows the number of
photons ($N$) that would be detected in IceCube for the case of
three active neutrinos only with either normal or inverted
hierarchy. For inverted hierarchy we show the results where we
disregard the shock effect (NS), consider only the effect of the
forward shock (FS) and take effect of both the forward and reverse
shocks (FRS). For normal hierarchy, the shock does not have any
effect and we have the results denoted as N3 in the figure for all
cases. The right hand panel shows the difference ($\Delta N$)
between the number of photons expected in presence of shock to the
number of photons in absence of shock effects. The observed photons
are split into 100ms time bins and the width of the lines reflect
the expected fluctuation in the photons detected. }
\end{figure}

\begin{figure}[h]
\begin{center}
\hglue -2.5cm
\includegraphics[width=17cm]{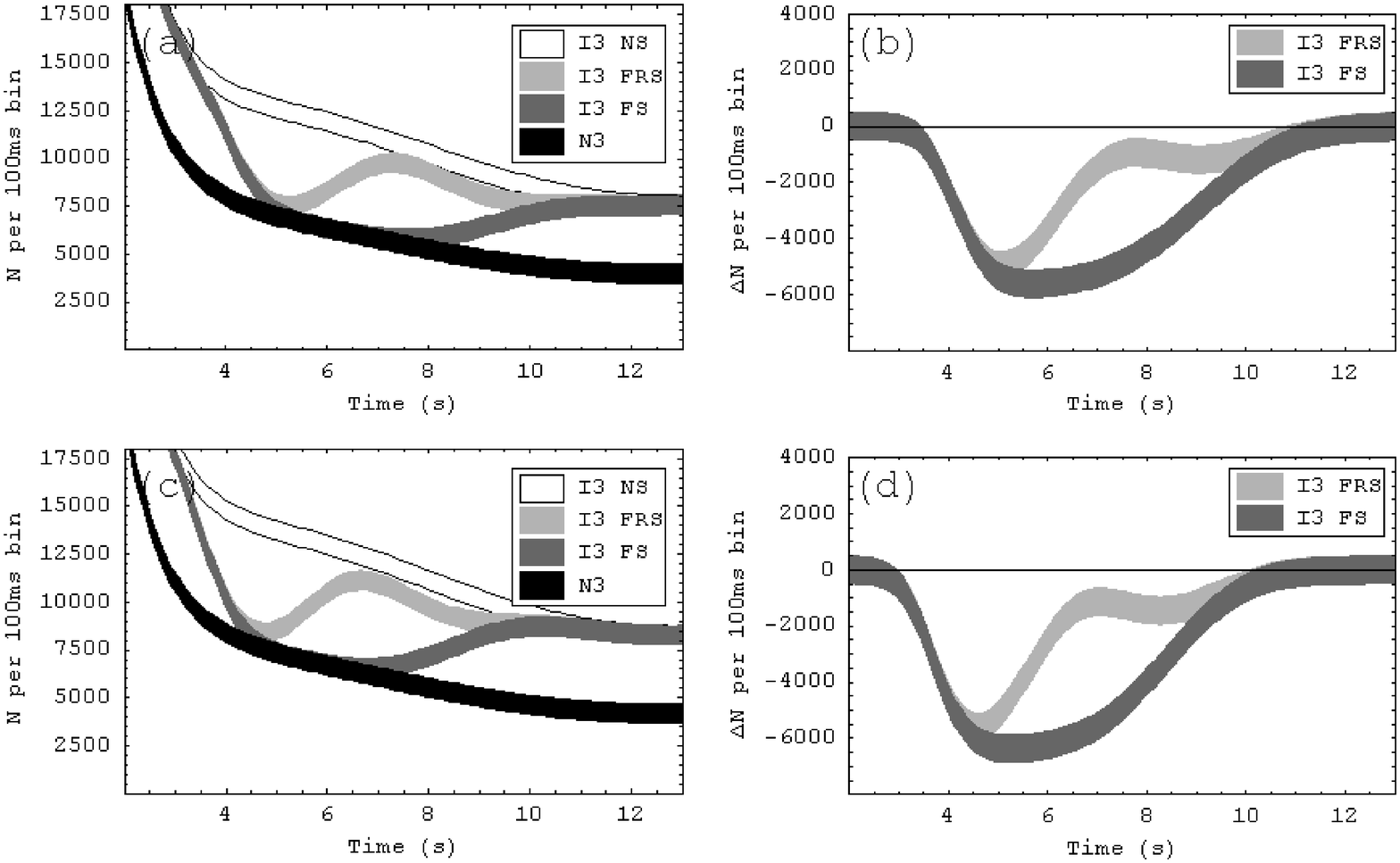}
\end{center}
\caption{Same as Fig. \ref{fig:3active} but with equal initial
energy spectra for the active antineutrinos and different
luminosities as discussed in the text. (a) and (b) are for the case
called G1 while (c) and (d) show the case G2.} \label{fig:3activeb}
\end{figure}

From Eq. (\ref{eq:flux1}) we see that if the initial flux
$F^0_\alpha(E)$ of the $\bar\nu_{e}$ and $\bar\nu_{x}$ are
identical, there will be no observable oscillation effect. In the LL
case the average energy of the $\nu_x$ is much larger than that of
the $\anue$. Since the detection cross-section increases
quadratically with anti-neutrino energy, flavor oscillations inside
the supernova are expected to enhance the anti-neutrino signal in
IceCube. In Fig. \ref{fig:3active} the number of photons that would
be detected by IceCube due to the supernova anti-neutrinos is shown
for the normal hierarchy and for the inverted hierarchy for the case
of no shock wave, a forward shock wave and a forward and reverse
shock. In the figure the width of the line represents the upper and
lower event number as discussed in the previous Section. Thus
signals much larger than this width should be observable.

Due to the energy dependence of the interaction cross section in the
detector the more energetic component will give the larger signal.
In the case of a normal hierarchy there are no resonant level
crossings for the antineutrinos and the $\bar{\nu}_{e}$ flux on
earth originated as the $\bar{\nu}_{e}$ flux at the neutrinosphere
(c.f. Table \ref{tab:jumpprob3}). Since in this case the resultant
neutrino flux has the least energetic component, it explains why the
normal hierarchy case in Fig. \ref{fig:3active} corresponds to the
lowest bound for the number of photons in the detector at all times.
For inverted hierarchy if we ignore the effect of shock waves, then
for large values of $\theta_{13}$ the transition is completely
adiabatic and the resultant $\bar\nu_{e}$ flux on earth originated
as the $\bar{\nu}_{x}$ flux at the neutrinosphere. Since in this
case the $\anue$ flux at the detector is made up entirely of the
most energetic component, the resulting number of photons provide
the upper bound on the expected number of photons in the detector.
The effect of the shock wave is to bring about abrupt changes in the
oscillation probability as discussed in the Section
\ref{sec:shocks}. In particular, we had seen that the transition
probability flips from being completely adiabatic to almost
non-adiabatic at the shock front. Therefore, including the effect of
the shock wave means that the $\bar\nu_{e}$ flux on earth at the
time when the shock front crosses the resonance region originated as
the $\bar{\nu}_{e}$ flux at the neutrinosphere, causing the
reduction seen in the signal at these times in Fig.
\ref{fig:3active}. We see that the structure in the signal for the
forward and reverse shock does show a single and double bump
structure respectively, conforming to the shape of the oscillation
probability discussed in Section \ref{sec:shocks}.
%Note that if the energy spectra of different flavours are indeed
%different, as discussed in \cite{3neutrinoanalysis},
%this double bump in the
%total events as a function of time
%is less prominent than bumps in the average energy and a
%detector capable of measuring the energy would be
%useful to determine the
%detailed structure of the shock waves.

In both the single and double shock cases the structure in the
signal above the photon background is clearly visible, demonstrating
that IceCube will indeed be capable of distinguishing between the
normal and inverted hierarchy for the case the initial energy
spectra of different flavours differ. However, as discussed in
Section \ref{energyspectra}, the inclusion of further scattering
processes within the supernovae is expected to have the effect of
reducing the difference in the average energy between the
$\bar\nu_{e}$ and $\bar\nu_{x}$ components. In this case the
structures in the signal just discussed and indeed the effect of
oscillations itself due to the energy differences between the
different (anti)neutrino flavors will not be present. However all is
not lost because the effect of the additional scattering processes,
while reducing the energy difference, increases the difference in
number flux of the neutrinos with the number flux of the
$\bar\nu_{e}$ becoming almost twice that of the $\bar\nu_{x}$
components during the accretion phase (cf. Table
\ref{tab:enlmodel}). To make this quantitative we have calculated
the number of photons expected in the IceCube detector for the cases
G1 and G2 of Table \ref{tab:enlmodel}. Since results for the
Garching simulations are not available over the full time period of
the supernova, we have used the average energy and total fluxes for
G1 and G2 from Table \ref{tab:enlmodel} but the time dependent fall
of the flux from the simulations of LL. The results are presented in
Fig. \ref{fig:3activeb}, where the upper panels are for the case G1
and lower ones for G2. The results we get are very similar to the
one we had for the LL case. The reason is that even though the ratio
of $\langle E^0_{\anue} \rangle/\langle E^0_{\nu_x} \rangle$ is
nearly 1 for the Garching simulations, the ratio of the fluxes
${\Phi^0_{\anue}}/{\Phi^0_{\nu_x}}$ is nearly half. This means that
in this case even though neutrino oscillation does not change the
energy of the resultant $\anue$ spectra emerging from the supernova,
it changes the total number flux, increasing the number of $\anue$
and hence producing an enhanced signal in the detector.

Note that the mere existence of structure due to the shock waves
provides the evidence for the inverted hierarchy. However as is
evident from Figs. \ref{fig:3active} and \ref{fig:3activeb} the
signal contains much more information, the sign of the effect giving
information about the relative importance of an energy difference or
luminosity difference between the $\bar\nu_{e}$ and $\bar\nu_{x}$
components. Also the shape of the signal provides seismological
information capable of determining much about the nature of the
supernova shock wave or waves. Clearly, given the sensitivity of the
signal to the initial spectra and luminosities, it is desirable to
have a full time dependent simulation of the luminosity and energy
spectrum of the Lawrence Livermore type but including all
significant neutrino interactions within the neutrinosphere.

%%%%%%%%%%%%%%%%%%%%%%%%%%%%%%%%%%%%%%%%%%%%%%%%%%
\subsubsection{Three Active and Two Sterile Neutrinos}
%%%%%%%%%%%%%%%%%%%%%%%%%%%%%%%%

\begin{table}[p]
\begin{center}
\begin{tabular}{|c|c|c|c|}
\hline Hierarchy & i & $P^{m}_{ei}$ & $P^{m}_{xi}$ \\ \hline\hline
N2+N3 & 1 & 1 & 0 \\
& 2 & 0 & $P_{25}P_{24}$ \\
& 3 & 0 & $%
P_{25}(1-P_{24})(1-P_{34})+(1-P_{25})(1-P_{35})P_{34}+P_{35}P_{34} $ \\
& 4 & 0 & $%
P_{25}(1-P_{24})P_{34}+(1-P_{25})(1-P_{35})(1-P_{34})+P_{35}(1-P_{34})$ \\
& 5 & 0 & $(1-P_{35})+(1-P_{25})P_{35}$ \\ \hline
N2+I3 & 1 & $P_{13}$ & $P_{35}P_{34}(1-P_{13})$ \\
& 2 & 0 & $%
P_{35}(1-P_{34})(1-P_{24})+(1-P_{35})(1-P_{25})P_{24}+P_{25}P_{24} $ \\
& 3 & $1-P_{13}$ & $P_{35}P_{34}P_{13}$ \\
& 4 & 0 & $%
P_{35}(1-P_{34})P_{24}+(1-P_{35})(1-P_{25})(1-P_{24})+P_{25}(1-P_{24})$ \\
& 5 & 0 & $(1-P_{25})+(1-P_{35})P_{25}$ \\ \hline
H2+N3 & 1 & $P_{14}$ & 0 \\
& 2 & 0 & $P_{25}$ \\
& 3 & 0 & $P_{35}+(1-P_{25})(1-P_{35})$ \\
& 4 & $1-P_{14}$ & 0 \\
& 5 & 0 & $(1-P_{35})+(1-P_{25})P_{35}$ \\ \hline
H2+I3 & 1 & $P_{14}P_{13}$ & $P_{35}(1-P_{13})$ \\
& 2 & 0 & $(1-P_{35})(1-P_{25})+P_{25}$ \\
& 3 & $(1-P_{13})P_{14}$ & $P_{35}P_{13}$ \\
& 4 & $(1-P_{14})$ & 0 \\
& 5 & 0 & $(1-P_{25})+(1-P_{35})P_{25}$ \\ \hline
I2+N3 & 1 & $P_{15}P_{14}$ & 0 \\
& 2 & 0 & 1 \\
& 3 & 0 & 1 \\
& 4 & $P_{15}(1-P_{14})$ & 0 \\
& 5 & $1-P_{15}$ & 0 \\ \hline
I2+I3 & 1 & $P_{15}P_{14}P_{13}$ & $1-P_{13}$ \\
& 2 & 0 & 1 \\
& 3 & $P_{15}P_{14}(1-P_{13})$ & $P_{13}$ \\
& 4 & $P_{15}(1-P_{14})$ & 0 \\
& 5 & $1-P_{15}$ & 0 \\
&  &  &  \\ \hline
\end{tabular}%
\label{tab:jumpprob}
\end{center}
\caption{The probabilities $P_{ei}^m$ and $P_{xi}^m$ for three
active neutrinos plus two sterile neutrinos, where
$P_{\protect\alpha i}^{m}$ is given by Eq. (\ref{eq:flipprob}) and
$P_{ij}$ is the flip probability at the resonance between the
$\protect\nu _{i}$ and $\protect\nu _{j}$ mass eigenstates.}
\label{flipprobthreeplustwo}
\end{table}

\begin{table}[tbp]
\begin{center}
\begin{tabular}{|c|c|c|}
\hline Hierarchy & $F_{no shock}$ & $F_{shock}$ \\ \hline
N2+N3 & $|U_{e1}|^{2}F^{0}_{e}$ & $P_{24}P_{25}|U_{e2}|^{2}F^{0}_{x}$ \\
N2+I3 & $|U_{e3}|^{2}F^{0}_{e}+(|U_{e4}|^{2}+|U_{e5}|^{2})F^{0}_{x}$ & $%
|U_{e1}|^2 P_{13}F_{e}^{0}+((P_{24}+P_{25})|U_{e2}|^{2}+P_{24}P_{25}
(|U_{e1}|^{2}-|U_{e2}|^{2}))F_{x}^{0}
$ \\
H2+N3 & $|U_{e4}|^{2}F^{0}_{e}+(|U_{e3}|^{2}+|U_{e5}|^{2})F^{0}_{x}$ & $%
P_{24}|U_{e1}|^{2}F^{0}_{e}+P_{25}|U_{e2}|^{2}F^{0}_{x}$ \\
H2+I3 & $|U_{e2}|^{2}F^{0}_{x}$ & $P_{25}|U_{e1}|^{2}F^{0}_{x}$ \\
I2+N3 & $|U_{e2}|^{2}F^{0}_{x}$ & $P_{24}P_{25}|U_{e1}|^{2}F^{0}_{e}$ \\
I2+I3 & $(|U_{e1}|^{2}+|U_{e2}|^{2})F^{0}_{x}$ &
$-P_{13}|U_{e1}|^{2}F^{0}_{x}$
\\ \hline
\end{tabular}%
\end{center}
\caption{\label{tab:jumpprobapprox}
The flux of neutrinos in the approximation that $%
|U_{e1}|^{2},|U_{e2}|^{2}>>|U_{e3}|^{2},|U_{e4}|^{2},|U_{e5}|^{2}$ and $%
P_{1i},P_{3i}\simeq P_{2i}$, where i=4 or 5.}
\label{flipprobsimplified}
\end{table}

We turn now to the case in which there are additional sterile
neutrinos capable of explaining the LSND anomaly. Once again
resonant effects due to the passage of shock waves through the
supernova can give significant information about the neutrino
mixing. In this case we do not have to rely on a detailed knowledge
of the luminosities and energy spectra of the active neutrinos
because oscillation to a sterile neutrino necessarily corresponds to
a reduction in the signal and therefore it's much easier to observe
oscillation and shock effects in this case.

\begin{figure}[h]
\begin{center}
\hglue -2.5cm
\includegraphics[width=11cm]{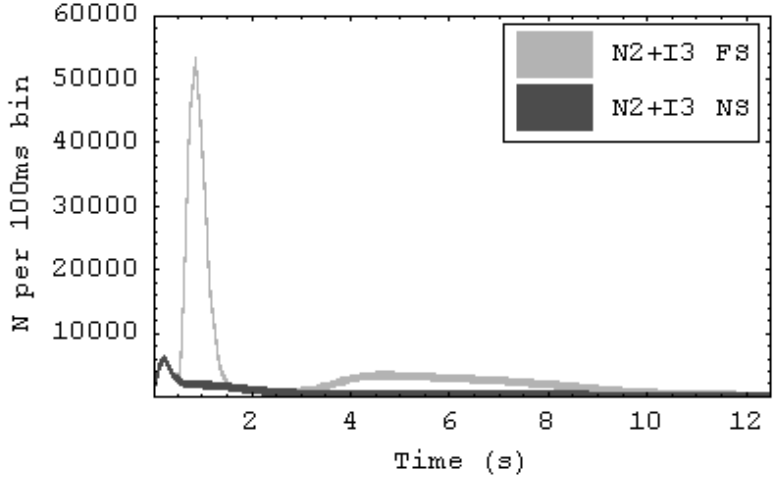}
\end{center}
\caption{The number of photons detected for a N2+I3 mass hierarchy
calculated for 100ms time bins with a forward shock (FS) and without
a shock (NS).} \label{fig:general}
\end{figure}

\begin{figure}[h]
\begin{center}
\hglue -2.5cm
\includegraphics[width=17cm]{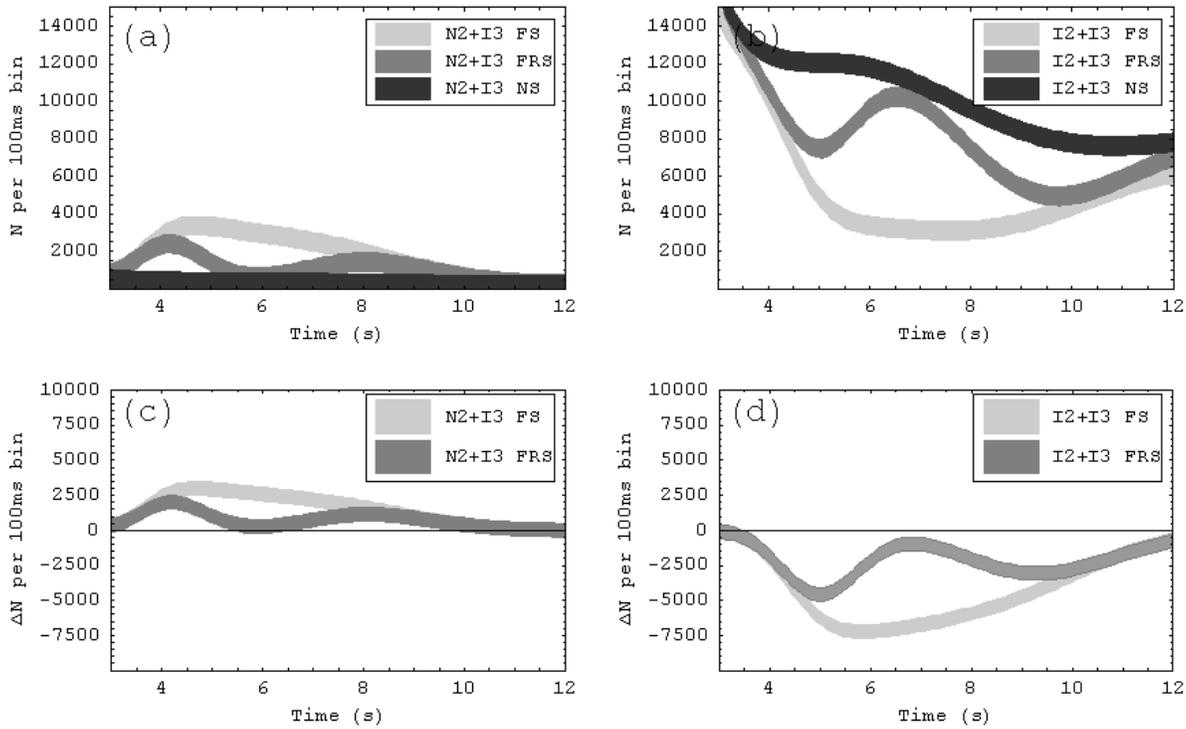}
\end{center}
\caption{The number of additional photons detected in 100ms time
bins, for the cases of a forward shock (FS), a forward and reverse
shock (FRS) and no shock (NS) for the mass hierarchies:(a) N2+I3,
(b) I2+I3. The lower panels show the corresponding difference in
between the number of photons expected due to the effect of the
shock for the mass hierarchies: (c) N2+I3, (d) I2+I3.}
\label{fig:resultlarge}
\end{figure}

\begin{figure}[p]
\begin{center}
\hglue -2.5cm
\includegraphics[width=17cm]{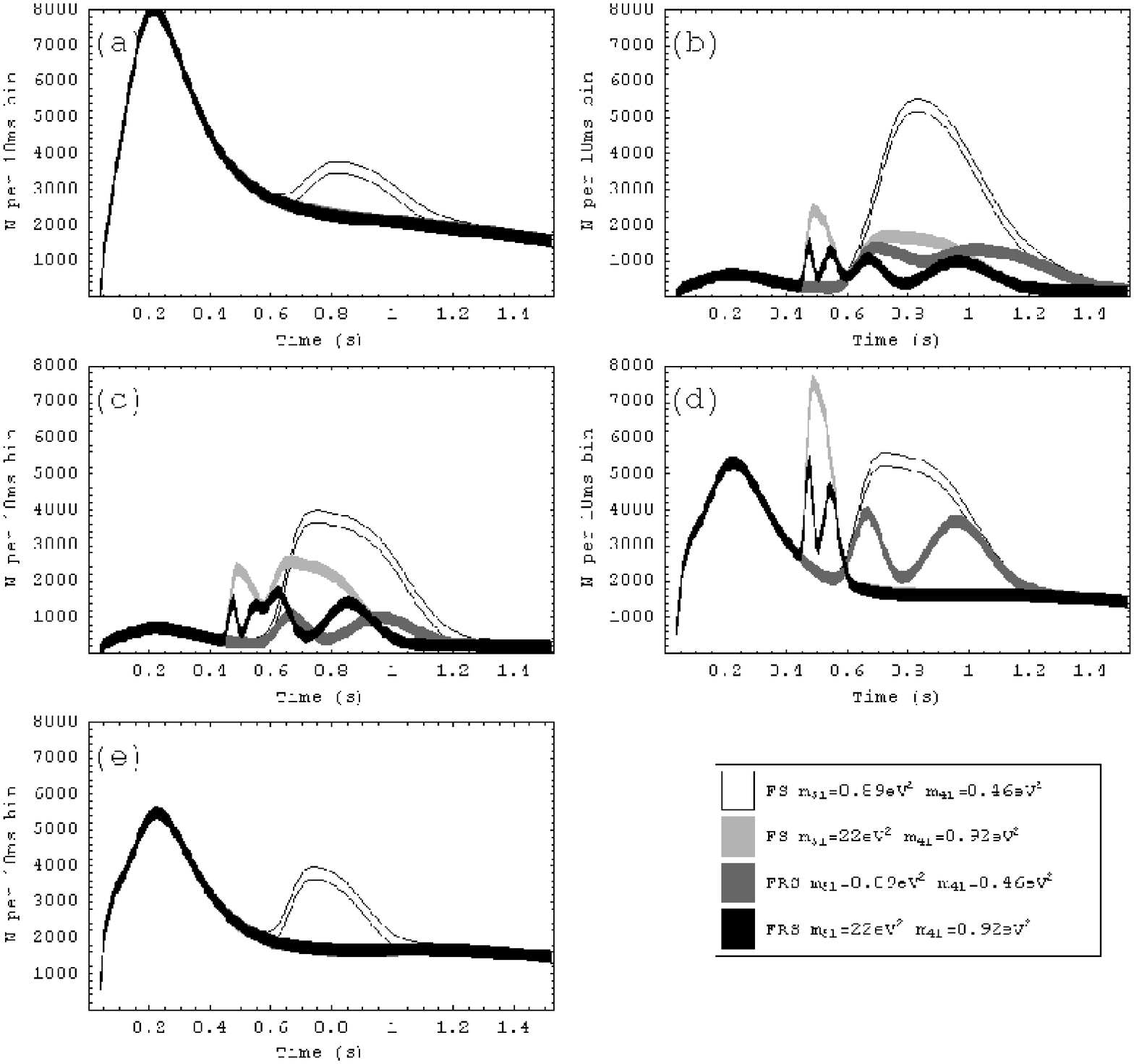}
\end{center}
\caption{The number of photons expected in 10ms time bins, due to a
forward shock (FS) and a forward and reverse shock (FRS), for the
mass hierarchies:(a) N2+N3, (b) N2+I3, (c) H2+N3, (d) H2+I3, (e)
I2+N3. We show the results for both the solutions given in Table
\ref{tab:accdata}. Note that the shock wave has no observable effect
for I2+I3.} \label{fig:resulthierarcya}
\end{figure}

\begin{figure}[p]
\begin{center}
\hglue -2.5cm
\includegraphics[width=17cm]{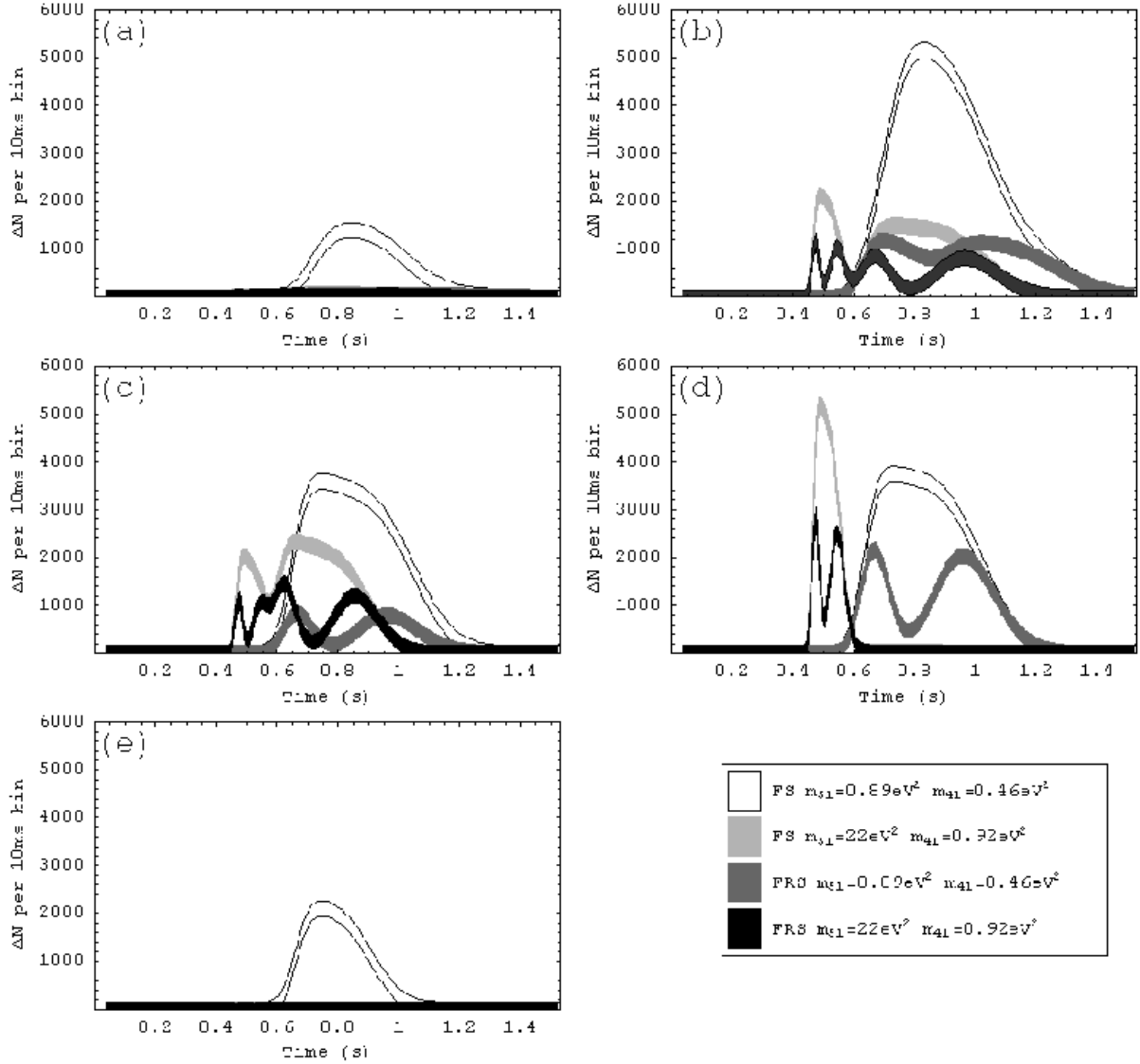}
\end{center}
\caption{The difference in the number of expected photons in 10ms
time bins due to the effect of a forward shock (FS) and a forward
and reverse shock (FRS), for the mass hierarchies: (a) N2+N3, (b)
N2+I3, (c) H2+N3, (d) H2+I3, (e) I2+N3.} \label{fig:resulthierarcyb}
\end{figure}

\begin{figure}[p]
\begin{center}
%\hglue -2.5cm
\includegraphics[width=17cm]{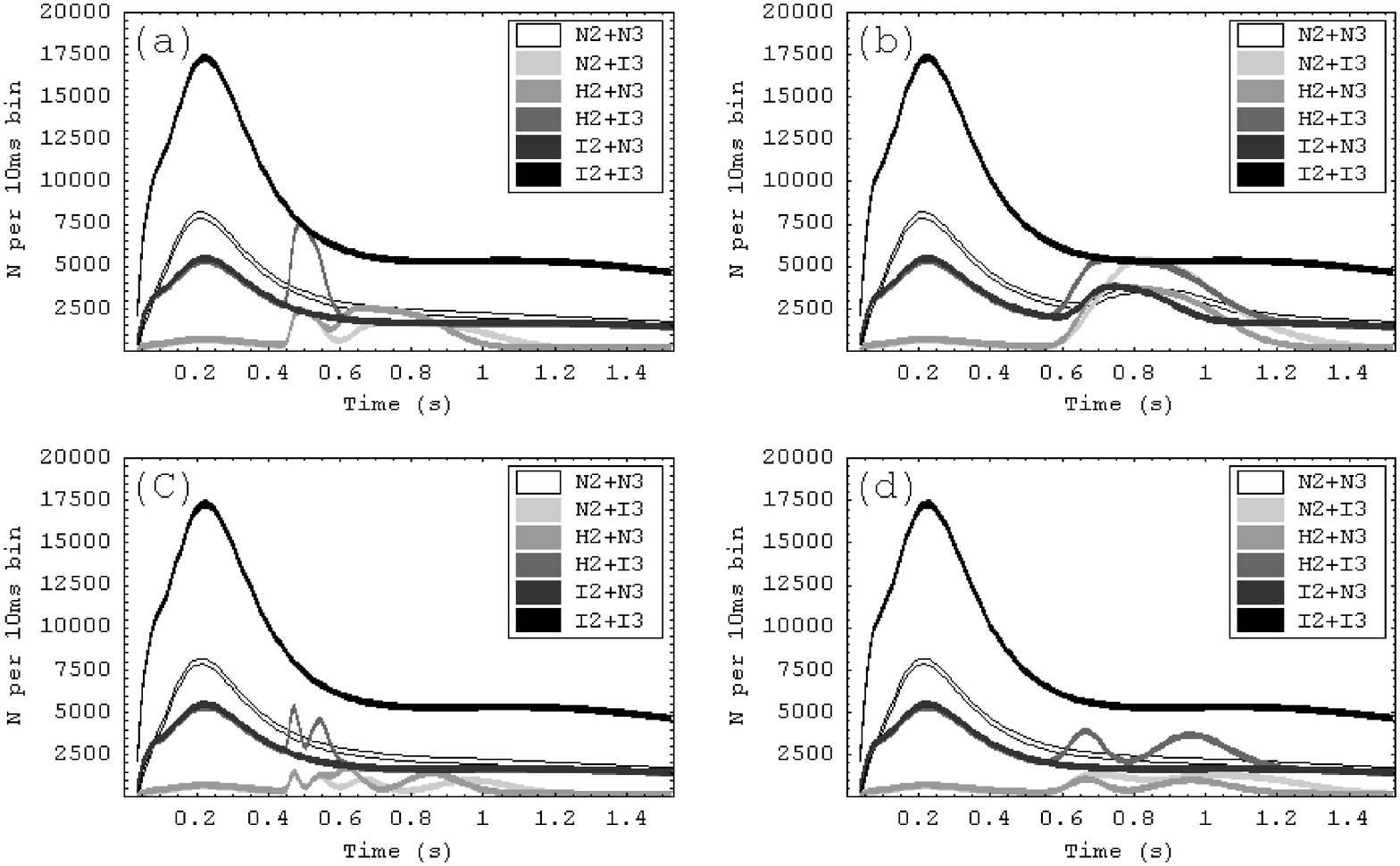}
\end{center}
\caption{The number of photons detected in 10ms time bins due to the
shockwave in various mass hierarchies, where (a) is for forward shock and $%
\Delta m^2_{51}=22eV^{2}$, $\Delta m^2_{41}=0.92eV^{2}$, (b) is for
forward shock and $\Delta m^2_{51}=0.89eV^{2}$, $\Delta
m^2_{41}=0.46eV^{2}$, (c) is for forward and reverse shock and
$\Delta m^2_{51}=22eV^{2}$, $\Delta m^2_{41}=0.92eV^{2}$, (d) is for
forward and reverse shock and $\Delta m^2_{51}=0.89eV^{2}$, $\Delta
m^2_{41}=0.46eV^{2}$.} \label{fig:resultshocka}
\end{figure}

\begin{figure}[p]
\begin{center}
\includegraphics[width=17cm]{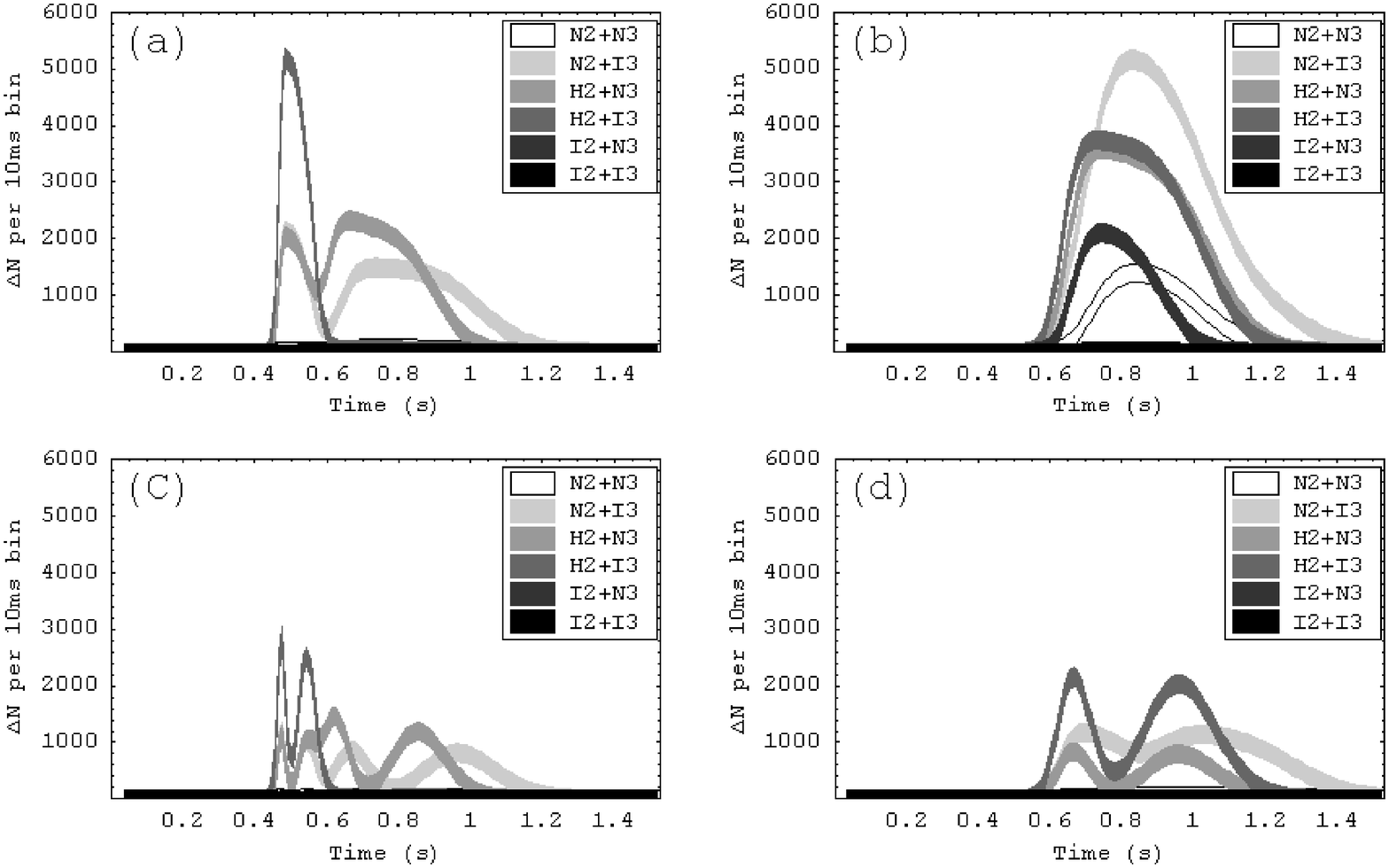}
\end{center}
\caption{The number of additional photons detected in 10ms time bins
due to the shockwave in various mass hierarchies, where (a) is for
forward shock and $\Delta m^2_{51}=22eV^{2}$, $\Delta
m^2_{41}=0.92eV^{2}$, (b) is for forward shock and $\Delta
m^2_{51}=0.89eV^{2}$, $\Delta m^2_{41}=0.46eV^{2}$, (c) is for
forward and reverse shock and $\Delta m^2_{51}=22eV^{2}$, $\Delta
m^2_{41}=0.92eV^{2}$, (d) is for forward and reverse shock and
$\Delta m^2_{51}=0.89eV^{2}$, $\Delta m^2_{41}=0.46eV^{2}$.}
\label{fig:resultshockb}
\end{figure}

\begin{figure}[p]
\begin{center}
\includegraphics[width=17cm]{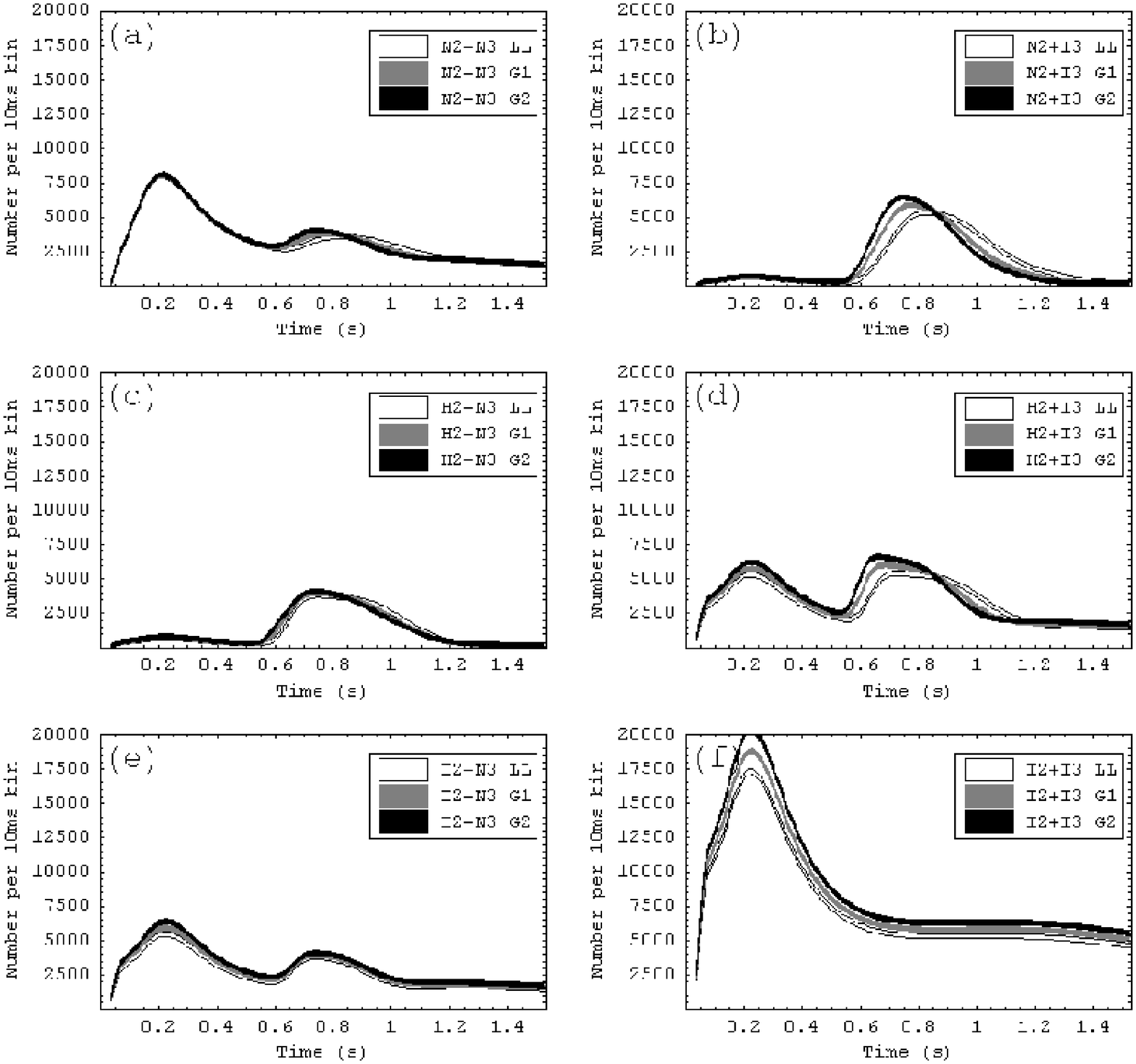}
\end{center}
\caption{\label{fig:raffeltsterile} Effect of changing the supernova
neutrino model on the resultant signal in the detector. The number
of photons expected in 10ms time bins, due to a forward shock and
$\Delta m^2_{51}=0.89eV^{2}$, $\Delta m^2_{41}=0.46eV^{2}$, for the
mass hierarchies: (a) N2+N3, (b) N2+I3, (c) H2+N3, (d) H2+I3, (e)
I2+N3 (f) I2+I3.}
\end{figure}

The analysis proceeds in a manner similar to that for the three
neutrino case. We start with the evolution of the mass squared of
the mass eigenstates for the different possible hierarchies as shown
in Fig. \ref{fig:3+2mass}. From this it is straightforward to
compute the $\bar\nu_{e}$ flux in the detector using Eqs.
(\ref{eq:fluxprob}) and (\ref{eq:flipprob}). Table
\ref{flipprobthreeplustwo} gives the probabilities $P_{ei}^m$ and
$P_{xi}^m$ for the different neutrino mass spectra.
This simplifies for the case considered here, $%
|U_{e1}|^{2},|U_{e2}|^{2}>>|U_{e3}|^{2},|U_{e4}|^{2},|U_{e5}|^{2}$
giving the results in Table \ref{flipprobsimplified}. The angles
$\theta_{14}$ and $\theta_{15}$ are taken from the best fit values
of table \ref{tab:accdata}. We take
$\theta_{34}=\theta_{35}=10^{-3}$; note that these are not
constrained from any data. Finally we take $\theta_{13}=0.142$,
within the bounds from equation \ref{eq:CHOOZ} and
$\theta_{24}=0.12$ and $\theta_{25}=10^{-3}$,chosen such that the
LSND mixing angle $\theta_{LSND}$ is equal to it's best fit value,
where
$\sin^{2}(2\theta_{LSND})=4(U_{e4}U_{\mu4}+U_{e5}U_{\mu5})^{2}$.
These angles are large such that the transition probabilities change
from being adiabatic to non-adiabatic as the shock wave passes
through each resonance. The signal is dominated by detection of
anti-neutrinos leaving the supernova in the first and second mass
eigenstate. In all but one case the mixing angles are sufficiently
small such that the multiple resonances that these anti-neutrinos
cross are far enough apart such that they do not overlap and
therefore each resonance can be considered as independent. The
exception is the combination of resonances corresponding to $\Delta
m^{2}_{43}$ and $\Delta m^{2}_{42}$, however numerical checks have
shown that the total transition probability is still well
represented by a product of the two flip probabilities. Using this
we have computed the number of photons that would be detected in
IceCube for a galactic supernova. Figs. \ref{fig:general},
\ref{fig:resultlarge}, \ref{fig:resulthierarcya},
\ref{fig:resulthierarcyb}, \ref{fig:resultshocka} and
\ref{fig:resultshockb} are for the LL case while Fig.
\ref{fig:raffeltsterile} compares the impact of the change in the
initial neutrino fluxes and spectra from LL to G1 and G2.

Fig. \ref{fig:general} shows the number of photons detected in time
bins of 100 ms, due to a supernova for neutrinos in a N2+I3 mass
hierarchy, with and without a shock wave. There are bumps in the
signal due to the propagation of the shock wave in the first 2s and
from 4s-11s. As has been discussed for the case of just three active
neutrinos, the latter structure is due to the propagation of the
shockwave through the resonant density corresponding to the
atmospheric mass splitting. The bumps in the first 1.5s is due to
the propagation of the shockwave through the resonant densities
corresponding to the sterile neutrinos.

We first concentrate on the $\ma$ driven ``atmospheric resonance''
to determine the changes due to the mixing with sterile neutrinos.
The effect of the shock wave is only relevant if the three active
neutrinos have an inverse hierarchy in their masses. This is evident
from Table \ref{tab:jumpprobapprox}, where the flip probability
$P_{13}$ appears only for cases with I3. For these cases we show in
Fig. \ref{fig:resultlarge} the number of photons detected on the
same time bins as that shown in Fig. \ref{fig:3active}. We do not
show the case for H2+I3 because, c.f. Table
\ref{tab:jumpprobapprox}, the shock wave only causes a change in the
signal if it also passes through a sterile resonance at the same
time. In theoretical simulations of the shockwave this does not
happen and therefore no \textquotedblleft bump\textquotedblright\
should be observed. From the Figs. \ref{fig:3active} and
\ref{fig:resultlarge} we see that the N2+I3 hierarchy gives an
increase in the signal compared to the case of only 3 active
neutrinos while the I2+I3 case shows a decrease. In both cases we
get a statistically significant signal.

Turning to the sterile resonance region we present in Figs. \ref%
{fig:resulthierarcya}, \ref{fig:resulthierarcyb}, \ref{fig:resultshocka} and %
\ref{fig:resultshockb} our numerical results for the sterile
resonances. The structure of the signal spans of the order of 0.1s
and therefore a detector needs to have a time resolution of at least
10ms to try to disentangle the information. The number of bumps is
determined by several factors. First, it is dependent on the mass
hierarchy which determines the number of resonances which the
neutrino passes through, (see Table \ref{flipprobthreeplustwo}). If
a neutrino passes through two sterile resonances it could create two
bumps; however if the resonances are close inside the supernova,
only one broader bump may be observed. If the shockwave has a
forward and reverse shock the double peak structure in the
probability can cause several bumps in the signal, as can clearly be
seen in Fig. \ref{fig:resulthierarcyb} (c) and (d). For the N2+I3
and H2+N3 mass
hierarchies, because $%
|U_{e1}|^{2},|U_{e2}|^{2}>>|U_{e3}|^{2},|U_{e4}|^{2},|U_{e5}|^{2},$
the number of events in the absence of a shock is an order of
magnitude smaller than that of other hierarchies. Therefore
observation of a large ratio of the height of the bump to the
background supernova signal would be characteristic of these mass
hierarchies. However in practice this could be misidentified if the
luminosity varied rapidly in time.

The effect of changing the average energy and total number flux of
the neutrinos is shown in Fig. \ref{fig:raffeltsterile}. The six
panels are for the six different mass spectra and we consider the
case where we have both forward and reverse shocks. We can see from
the figure that changing the supernova neutrino model from LL to G1
or G2 does not wipe out any of the interesting structures in the
signal.

%%%%%%%%%%%%%%%%%%%%%%%%%%%%%%%%%%%%%%
\section{Summary and Conclusions}
%%%%%%%%%%%%%%%%%%%%%%%%%%%%%%%%%

Our analysis of the neutrino signals in the IceCube detector that
would result from a \textquotedblleft standard\textquotedblright\
supernova at 10 kpc from earth has shown that oscillations to
sterile neutrinos will be easily detectable for a wide variety of
neutrino spectra. As may be seen from the figures, for 10 msec time
bins, the event rate is typically very large and supernovae as far
as about 30 kpc from earth will still give observable effects. The
reason the signals are so distinctive is a result of the combination
of resonant conversion within the supernova with the time dependent
structure resulting from the propagation of a shock wave through the
supernova. Together this gives a characteristic time dependent
signal that is readily observable with a detector time resolution
better than 10 msec. Since oscillation to sterile neutrinos
generates a disappearance event the results are relatively
insensitive to uncertainties in the initial energy spectra and
luminosities.

The signal coming from resonant effects involving just three
antineutrinos with an inverted mass hierarchy is much more sensitive
to the initial spectra and luminosities because the signal vanishes
if both are degenerate for the three flavours. For the case the
energy spectra differ significantly, as in the Lawrence Livermore
simulation, the resulting signal is again clearly visible in
IceCube. However if the energy spectra are degenerate, as recent
studies seem to indicate, then one must rely on differences in the
initial luminosities. To date no full simulation has been performed
over the time period of interest but extrapolating the recent
simulations, which extend up to 750 ms, to later times suggests that
even in this case the resonant conversion in the presence of a shock
wave would be clearly visible despite a reduction of about a factor
of two in the signal. To improve our confidence in this result it is
clearly of importance to have a reliable simulation of neutrino
propagation through the supernova over a 10 second period including
all the significant antineutrino processes.

Observation of sharp time dependent neutrino oscillation signals
will also provide a probe of the propagation of the shock wave
through the supernova. As our simulations show it is possible that
one can see both the forward and reverse shocks and the structure of
the signal will give information about their spatial structure. At
present our knowledge about the nature of the shock is quite
primitive and clearly better simulations of the shock properties
will improve our ability to interpret the signals. A very recent
paper\cite{turbulence} has pointed out that turbulence after the
shock can significantly affect the signals, broadening the
structure, and clearly it is also of importance to study this in
more detail.

In summary we have found that the detailed time structure of the
neutrinos coming from supernovae can provide information, not
accessible by other means, both on the mass spectrum and mixing
angles of neutrinos and on supernovae seismology. This applies to
those neutrino mass hierarchies and mixing angles, both for three
active neutrinos and for a combination of active and sterile
neutrinos, which give rise to resonant transitions within the
supernovae. For the case involving sterile neutrinos our analysis
has concentrated on the parameter space which can explain all
neutrino oscillation phenomena including the LSND measurement.
However, even if the LSND result should not be confirmed, the
sensitivity of experiments such as IceCube is such that they will
still be able to probe for a sterile neutrino component not
otherwise observable.

\noindent{\large \bf Acknowledgment}

We wish to thank Subir Sarkar for discussions. This work was
partially supported by the EC 6th Framework Programme
MRTN-CT-2004-503369, and a PPARC studentship.

%\appendix

\section*{Appendix: Detector properties}
\label{app:det}

The following analysis is the same as that presented by Dighe, Keil
and Raffelt\cite{3nusnactive}. For simplicity only events detected
by the $\bar{\nu}_{e}p\rightarrow ne^{+}$ are considered. Each
$e^{+}$ emits Cerenkov light according to

\begin{equation}
\frac{d^{2}N_{\gamma}}{dxd\lambda}=\frac{2\pi\alpha\sin^{2}\theta}{%
\lambda^{2}}
\end{equation}
where $N_{\gamma}$ is the number of emitted photons $x$ is the
distance traveled by the $e^{+}$, $\lambda$ is the wavelength of the
emitted photons, $\alpha$ is the fine structure constant and
$\theta$ is the angle of the
emitted photons to the direction of the $e^{+}$ such that $%
\cos\theta=(n\beta)^{-1}$, where n is the refractive index of the ice $%
(n_{ice}=1.31)$ and $\beta$ is the ratio of the speed of the
positron in the medium to that in a vacuum $(\beta_{ice}\sim1)$.
Integrating over the
observable wavelengths, taking the mean free path of a positron to be $%
12\,cm $ for an energy of $20\,MeV$ and assuming the two are
proportional the number of photons emitted between observable
wavelengths is:

\begin{equation}
\frac{N_{\gamma}}{E_{\bar{\nu}_{e}}}=191MeV^{-1}f_{Ch}
\end{equation}
where $E_{\bar{\nu}_{e}}$ is the energy of the $\bar{\nu}_{e}$ and
$f_{Ch}$ is a fudge factor. The number of detected photons per
optical module is:

\begin{equation}
N_{det}=\frac{47.75}{MeV m^{-3}}
f_{Ch}f_{abs}f_{OM}~\rho\frac{1}{4\pi D^{2}}
\int^{t_{max}}_{t_{min}}dt\int^{%
\infty}_{0}dE\sigma(E) F(E,t) E
\end{equation}
\begin{equation}
f_{OM}=\frac{Q}{0.20}\frac{A_{cat}}{250cm^{2}}\frac{\Omega_{acc}}{2\pi}
\end{equation}

\begin{equation}
f_{abs}=R_{abs}/100m
\end{equation}
where $\rho=6.18\times10^{25}$ m$^{-3}$ is the density of targets in
ice, $\sigma$ (in cm$^2$) is the cross section for
$\bar{\nu}_{e}p\rightarrow ne^{+}$, $E$ (in MeV) is energy, D
(cm$^{2}$) is the distance to the supernova, t (in s) is time, F (in
MeV$^{-1}$ s$^{-1}$) is the flux of anti-electron neutrinos, $Q$ is
the average quantum efficiency of the detector, $A_{cat}$ is the
effective photo cathode detection area, $\Omega_{acc}$ is the
angular acceptance range and $R_{abs}$ is the absorption length of
photons by the ice.

\end{document}